# RADIAL VELOCITIES OF 41 *KEPLER* ECLIPSING BINARIES

Rachel A. Matson[1,3], Douglas R. Gies, Zhao Guo[2,3]

*Center for High Angular Resolution Astronomy and Department of Physics and Astronomy, Georgia State University, P.O. Box 5060, Atlanta, GA 30302-5060, USA*

Stephen J. Williams[3]

*Department of Physics, University of Crete, P.O. Box 2208, Heraklion, Greece GR-71003*

[1]NASA Ames Research Center, Moffett Field, CA 94035, USA

[2]Copernicus Astronomical Centre, Bartycka 18, 00-716 Warsaw, Poland

[3]Visiting astronomer, Kitt Peak National Observatory, National Optical Astronomy Observatory, which is operated by the Association of Universities for Research in Astronomy (AURA) under a cooperative agreement with the National Science Foundation.

## ABSTRACT

Eclipsing binaries are vital for directly determining stellar parameters without reliance on models or scaling relations. Spectroscopically derived parameters of detached and semi-detached binaries allow us to determine component masses that can inform theories of stellar and binary evolution. Here we present moderate resolution ground-based spectra of stars in close binary systems with and without (detected) tertiary companions observed by NASA's *Kepler* mission and analyzed for eclipse timing variations. We obtain radial velocities and spectroscopic orbits for five single-lined and 35 double-lined systems, and confirm one false positive eclipsing binary. For the double-lined spectroscopic binaries we also determine individual component masses and examine the mass ratio $M_2/M_1$ distribution, which is dominated by binaries with like-mass pairs and semi-detached classical Algol systems that have undergone mass transfer. Finally, we constrain the mass of the tertiary component for five double-lined binaries with previously detected companions.

## 1. INTRODUCTION

Binary stars are ubiquitous throughout the galaxy and an important source of astrophysical parameters. Photometric and spectroscopic studies of eclipsing binaries, in particular, reveal fundamental stellar parameters such as masses and radii that inform our understanding of stars and constrain stellar evolutionary models. The frequency of binaries and their properties also serve as testbeds for star formation theories, essential for constraining current theoretical models of stellar and planetary formation alike. Known statistics of the field population for solar-type stars indicate at least 40% are binaries, with ∼12% in higher order multiples (Raghavan et al. 2010). Observational evidence, however, has shown that many such binaries are in fact triples (Eggleton et al. 2007; Tokovinin 2014a,b), especially short-period binaries with separations comparable to the stellar radii (Tokovinin et al. 2006). This prevalence of tertiary companions orbiting close binaries has strong implications for star formation mechanisms because the protostellar radii would be too large to fit inside their present day orbits (Rappaport et al. 2013). Theoretical studies, however, suggest that the presence of a third star can cause large eccentricity excitations to the inner orbit causing tidal forces to shrink and circularize the inner orbit via the eccentric Kozai-Lidov mechanism (Fabrycky & Tremaine 2007; Naoz 2016). This mechanism has also been proposed for planet migration, specifically to explain the presence of hot Jupiters with eccentric and misaligned orbits, as distant stellar or planetary companions with highly inclined orbits can perturb the planetary orbit and cause it to decay (Naoz et al. 2012).

rachel.a.matson@nasa.gov, gies@chara.gsu.edu, guo@camk.edu.pl

williams@physics.uoc.gr

One of several methods to find such triple-star systems involves long-term monitoring of binary eclipses for periodic perturbations caused by the presence of a third star. The nearly continuous photometry of over $150,000$ stars and more than 2000 eclipsing binaries (Kirk et al. 2016) collected by NASA's *Kepler* mission (Borucki et al. 2010) created an ideal data set for identifying eclipse timing variations, and has resulted in the discovery of hundreds of triple star candidates (Gies et al. 2012; Rappaport et al. 2013; Conroy et al. 2014; Gies et al. 2015; Borkovits et al. 2016).

In Gies et al. (2012, 2015) we reported eclipse timing variations for a subset of 41 eclipsing binaries chosen to optimize the chances of discovery of a third body in the system and enable follow-up ground-based spectroscopy. In total, we identified seven probable triple systems and seven additional systems that may be triples with orbits longer than the *Kepler* baseline (Gies et al. 2015). Subsequently, we have completed a large set of spectroscopic observations of this sample in order to determine spectroscopic orbits, estimate stellar properties, compare with evolutionary codes (Matson et al. 2016), and explore pulsational properties (Guo et al. 2016, 2017a,b) of the component stars.

Of the 41 eclipsing binaries selected for eclipse timing analysis via *Kepler*, approximately two-thirds were reported only recently to be eclipsing based on automated variability surveys such as the Hungarian-made Automated Telescope Network (HATnet), whose goal is to detect transiting extrasolar planets using small-aperture robotic telescopes (Hartman et al. 2004), and the All Sky Automated Survey (ASAS) which monitors $V$-band variability among stars brighter than 14th magnitude (Pigulski et al. 2009). Most of the remaining binaries have been known since prior epochs, but typically have little more than times of eclipse minima and orbital ephemerides published.

To characterize further this set of eclipsing binaries and derive spectroscopic orbital elements we collected an average of 11 ground-based optical spectra per binary. Ideally, when measuring radial velocities, high resolution spectra and complete phase coverage of the orbit are desired. However, moderate resolution ($R = \lambda/\delta\lambda \approx 6000$) optical spectra in the wavelength range $3930 - 4600$Å provided a high density of astrophysically important atomic lines and molecular bands (traditionally used for stellar classification) that allowed us to derive accurate radial velocities of intermediate-mass ($\sim 1 - 5\ M_\odot$) stars. In addition, the ephemerides determined in the eclipse timing analysis (Gies et al. 2015) enabled us to concentrate our observations during velocity extrema to best constrain the spectroscopic orbits with a modest number of spectra.

We discuss our observations in Section 2, followed by the determination of radial velocities and orbital parameters in Section 3. Discussion of the radial velocity results, mass ratio trends, and suspected triple systems is given in Section 4. Finally, a brief summary of our results is given in Section 5.

## 2. OBSERVATIONS

Spectra for all 41 eclipsing binaries were obtained over the course of six observing runs between 2010 June and 2013 August at Kitt Peak National Observatory (KPNO) with the 4 m Mayall telescope and R-C Spectrograph. Using the BL380 grating (1200 grooves mm$^{-1}$) in second order provided wavelength coverage of $3930 - 4600$Å with an average resolving power of $R = \lambda/\delta\lambda \approx 6200$. For wavelength calibration purposes, spectra of HeNeAr comparison lamps were taken either immediately before or after each science exposure, and numerous bias and flat-field spectra were taken each night.

Additional observations for sixteen of the brighter systems (*Kepler* magnitude, $K_p \lesssim 12$) were made at the Anderson Mesa Station of Lowell Observatory between 2010 July and 2012 November. The 1.8 m Perkins telescope and the DeVeny Spectrograph were used along with a 2160 grooves mm$^{-1}$ grating to obtain a resolving power of $R = \lambda/\delta\lambda \approx 6000$ over the wavelength range $4000 - 4530$Å. Calibration exposures with HgNeArCd Pen-Ray lamps were taken before or after each exposure while bias and flat-field spectra were taken nightly.

Ten of the binaries were also observed at the Dominion Astrophysical Observatory (DAO) 1.8 m Plaskett telescope in 2010 July. The Cassegrain Spectrograph was used with the 1200B grating in first order to obtain wavelength coverage from 4260 − 4600Å and an average resolving power of $R = \lambda/\delta\lambda \approx 4200$. Bias and flat-field exposures were taken nightly and FeAr comparison lamp spectra were taken immediately before or after each science exposure for wavelength calibration. A summary of the observations and spectral characteristics for all three setups is provided in Table 1.

**Table 1**: Spectroscopic Observations

| Observatory | Wavelength Range (Å) | Average Resolving Power ($\lambda/\delta\lambda$) | Average S/N | Number of Spectra |
|---|---|---|---|---|
| KPNO | 3930 – 4600 | 6200 | 100 | 367 |
| Lowell | 4000 – 4530 | 6000 | 40 | 48 |
| DAO | 4260 – 4600 | 4200 | 30 | 39 |

**Table 2**: Standard Velocity Stars

| Star | $T_{eff}$ (K) | log g (cgs) | $V_r$ (km $s^{-1}$) | Sources |
|---|---|---|---|---|
| HD 37160 | 4668 | 2.46 | 99.29 | Cenarro et al. 2007; Massarotti et al. 2008 |
| HD 82106 | 4868 | 4.80 | 29.84 | Nidever et al. 2002; Valenti & Fischer 2005 |
| HD 102870 | 6109 | 4.20 | 4.45 | Nidever et al. 2002; Cenarro et al. 2007 |
| HD 144579 | 5395 | 4.75 | -59.43 | Nidever et al. 2002; Luck & Heiter 2006 |
| HD 187691 | 6107 | 4.30 | -0.15 | Cenarro et al. 2007; Molenda-Zakowicz et al. 2007 |
| HD 194071 | 5486 | 2.70 | -9.43 | Latham & Stefanik 1992; Gray 2008 |
| HD 213947 | 4973 | 2.10 | 16.58 | Famaey et al. 2005; Gray 2008 |

All spectra were reduced and extracted using standard IRAF[1] routines. Wavelength calibration was performed using IRAF and the corresponding comparison lamp spectra for KPNO and DAO, while spectra from Lowell required observations of standard velocity stars to aid in the determination of the dispersion solution as the comparison lamps have insufficient lines in our wavelength regime. More details of the Lowell wavelength calibration can be found in Matson et al. (2016), however, we briefly summarize the process here. The comparison lamp exposures from Lowell were used to make an initial fit of wavelength to pixel number, which was augmented by cross-correlating the observed standard velocity stars with appropriate UVBLUE[2] models (Rodríguez-Merino et al. 2005) to get mean pixel and wavelength values in 40 sub-regions across the spectrum. These values were then fit with a cubic polynomial, which was combined with individual pixel shifts determined from the comparison lamp spectra to derive dispersion corrections for each science spectrum. Properties of the standard velocity stars used to calibrate Lowell spectra are shown in Table 2. After wavelength calibration, all spectra were rectified to a unit continuum and transformed to a common heliocentric wavelength grid of 1733 spectral steps in $\log\lambda$ increments equivalent to Doppler shift steps of 26.2 km $s^{-1}$ over the range 3950 to 4600Å. Six reduced and transformed spectra are shown in Figure 1, demonstrating the range of spectral types (B−G) in the sample and highlighting the changing spectral features visible in our wavelength range.

## 3. SPECTRAL ANALYSIS

### 3.1. *Radial Velocities*

To measure radial velocities we use a two-dimensional cross-correlation scheme, employing two templates to determine the velocity separation of the secondary component relative to the primary and the absolute velocity of the primary, based on the method used in PROCOR by R.W. Lyons (see Gies & Bolton 1986). Model templates for each star were selected from the UVBLUE grid of model spectra based on LTE calculations using the ATLAS9 and SYNTHE codes of R. L. Kurucz (Rodríguez-Merino et al. 2005). Templates for the primary and secondary were selected based on temperatures determined via the *Kepler* Input Catalog and temperature ratio as derived by Slawson et al. (2011) or using spectral energy distribution (SED) fits by Armstrong et al. (2014). The temperatures that best matched the observed spectral type and preliminary mass estimates were adopted and are given in Table 3. In a few cases the temperature ratio determined from the light curve and used as a prior in Armstrong et al. (2014) was used to provide a more reasonable temperature for the secondary. Detailed analyses of five systems (KIC 4544587, 5738698, 8262223, 9851944, and 10661783) include previously derived temperatures via spectroscopy and light curve analysis that we adopt instead. Gravities ($\log g$), projected rotational velocities ($v \sin i$), and initial estimates of the relative flux contribution of each star were calculated based on the temperatures, assuming main sequence stars and solar metallicity. Each model spectrum was then rebinned onto the observed wavelength grid (3950 − 4600Å) and convolved with functions for the projected rotational velocity and instrumental broadening. The adopted temperatures and monochromatic flux ratios (at 4275 Å) for each system are shown in Table 3 along with the *Kepler* Input Catalog (KIC) number, alternate object names, and *Kepler* magnitude ($K_p$).

[1] IRAF is distributed by the National Optical Astronomy Observatory, which is operated by the Association of Universities for Research in Astronomy (AURA), Inc., under cooperative agreement with the National Science Foundation.

[2] http://www.inaoep.mx/~modelos/uvblue/download.html

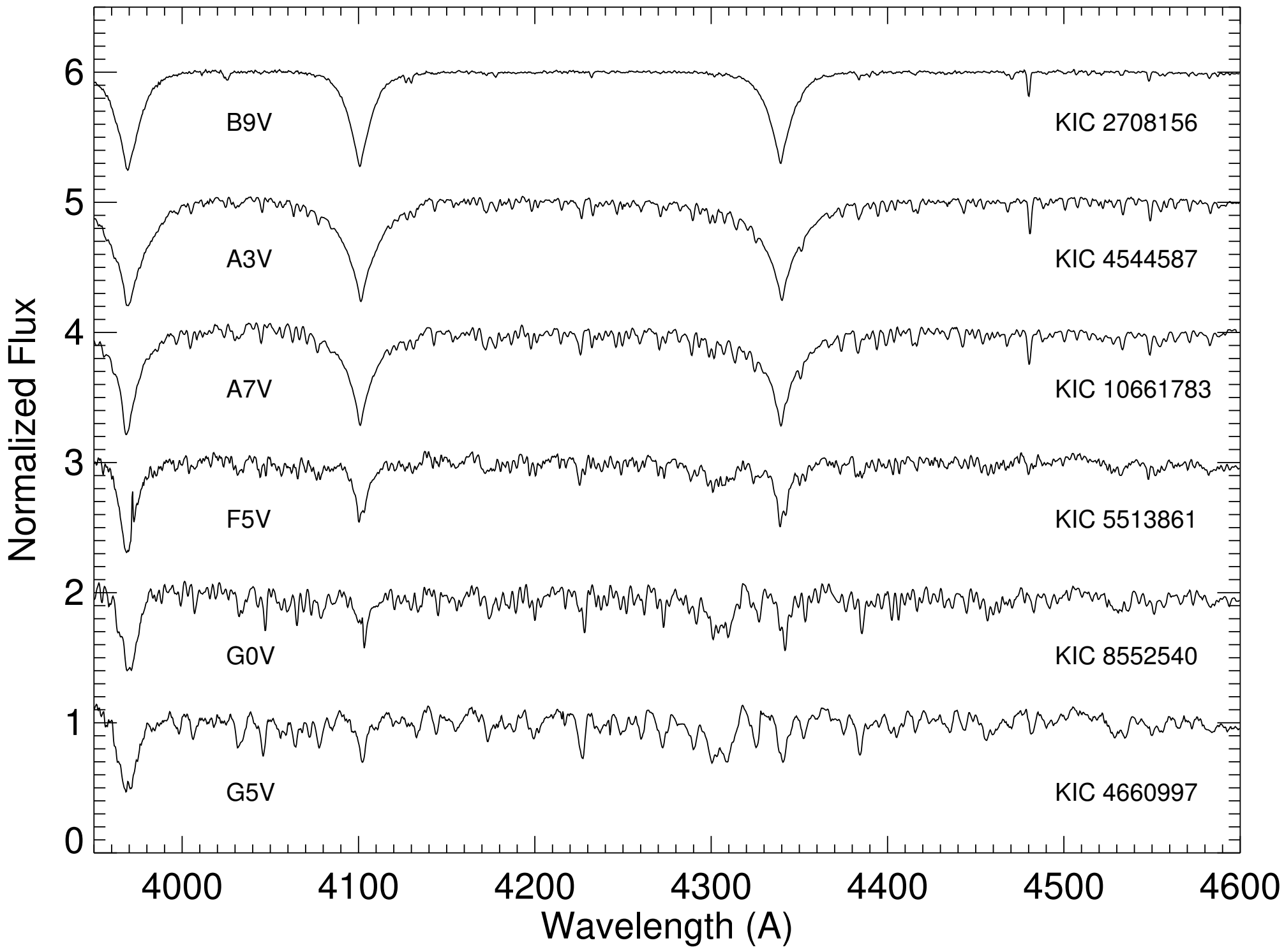

**Figure 1**: Representative spectra of six systems (offset for clarity) and their spectral types, including the hottest and coolest primary stars in our sample. The weakening of the hydrogen Balmer lines, H$\gamma$ (4340Å), H$\delta$ (4102Å), and H$\epsilon$ (3970Å), with decreasing temperature is evident, while the metal lines strengthen and the molecular G-band near 4300Å develops.

Using the times of observation and orbital elements estimated from the period and epoch of primary eclipse in Gies et al. (2015) and the temperatures and inclinations of Slawson et al. (2011), we predicted radial velocities for each observation to determine trial velocity separations for the primary and secondary spectral components (see Matson et al. 2016). These trial separations were used to make a series of composite model spectra that were cross-correlated with the observed spectrum. We then plotted the cross-correlation maxima versus trial separation and fitted the peak position to obtain the best-fit velocity separation of the primary and secondary. A final cross-correlation was performed using models constructed with this separation to get the absolute velocity of the primary, and by extension the secondary, for each spectrum. The uncertainties in the resulting velocities were estimated from maximum-likelihood principles using the method of Zucker (2003). Because the radial velocity measurements are derived from different instruments, we might also expect systematic differences in the velocities from KPNO, Lowell, and DAO. However, any such differences appear to be below our measurement sensitivities as we see no visible trends or offsets in the radial velocities or orbital solutions from different instruments. More careful examination of this issue for KIC 5738698 (Matson et al. 2016) and KIC 9851944 (Guo et al. 2016) similarly revealed no offsets in the data.

As broad hydrogen (and diffuse helium) lines are usually omitted when determining radial velocities from individual lines because blending effects can cause the line centers to appear displaced in wavelength (Petrie et al. 1967), we chose to omit the Balmer lines (H$\epsilon$, H$\delta$, H$\gamma$) from the cross-correlation. This was accomplished by blanking out the relevant pixels in each spectrum and using a Tukey (tapered cosine) window to smooth the edges and minimize systematic offsets in the derived radial velocities. To further aid in the accurate determination of radial velocities, we used the estimated velocity separation to guide which peak was chosen in the event of multiple peaks in the cross-correlation maxima versus separation plot. When multiple peaks were blended together, we performed a local rectification of

**Table 3**: Eclipsing Binary System Parameters

| KIC | Other ID | $K_p$ (mag) | Template $T_1$ (K) | Template $T_2$ (K) | Flux ratio[a] ($f_2/f_1$) | Source |
|---|---|---|---|---|---|---|
| 2305372 | 2MASS J19275768+3740219 | 13.821 | 6208 | 4097[b] | 0.01 | A |
| 2708156 | UZ Lyr | 10.672 | 11061 | 5671 | 0.01 | S |
| 3241619 | 2MASS J19322278+3821405 | 12.524 | 5715 | 4285 | $< 0.01$ | A |
| 3327980 | 2MASS J19084227+3826005 | 12.119 | 7321 | 6424 | 0.26 | S |
| 3440230 | 2MASS J19215310+3831428 | 13.636 | 8300 | 4897[b] | 0.05 | spec |
| 4544587 | TYC 3124-1348-1 | 10.801 | 8600 | 7750 | 0.66 | H |
| 4574310 | 2MASS J19395847+3938341 | 13.242 | 7153 | 4077 | 0.05 | A |
| 4660997 | V1130 Cyg | 12.317 | 5587 | 4215[b] | 0.05 | A |
| 4665989 | 2MASS J19390335+3945102 | 13.016 | 7559 | 6846 | 0.12 | A |
| 4678873 | TYC 3140-587-1 | 12.725 | 7496 | 5698 | 0.01 | S |
| 4848423 | KOI-3560/TYC 3140-2904-1 | 11.825 | 6239 | 6176 | 0.83 | A |
| 4851217 | HD 225524 | 11.108 | 7022 | 6804 | 2.00 | A |
| 5444392 | TYC 3138-829-1 | 11.378 | 5965 | 5726 | 0.86 | A |
| 5513861 | TYC 3123-2012-1 | 11.638 | 6479 | 6411 | 0.59 | A |
| 5621294 | 2MASS J19285262+4053359 | 13.613 | 8425 | 5560[b] | 0.03 | S |
| 5738698 | TYC 3141-1400-1 | 11.941 | 6792 | 6773 | 0.82[c] | M |
| 6206751 | 2MASS J19293751+4130469 | 12.142 | 6965 | 4885 | 0.02 | S |
| 7368103 | 2MASS J19333970+4255021 | 13.419 | 7838 | 5212[b] | 0.01 | S |
| 8196180 | 2MASS J20023258+4403122 | 12.814 | 7114 | 5934 | 0.04 | S |
| 8262223 | TYC 3162-1562-1 | 12.146 | 9128 | 6849 | 0.12 | G2 |
| 8552540 | V2277 Cyg | 10.292 | 5948 | 5252[b] | 0.17 | A |
| 8553788 | 2MASS J19174291+4438290 | 12.691 | 8045 | 5328 | 0.02 | S |
| 8823397 | 2MASS J19342636+4501070 | 13.249 | 8540 | 5724 | 0.07 | S |
| 9159301 | TYC 3556-2697-1 | 12.146 | 7959 | 4209 | 0.02 | S |
| 9357275 | 2MASS J19484858+4550595 | 12.186 | 7545 | 5580 | 0.01 | S |
| 9402652 | V2281 Cyg | 11.823 | 6641 | 6587[b] | 1.18 | S |
| 9592855 | 2MASS J19350483+4614117 | 12.216 | 7290 | 7087 | 0.64 | S |
| 9602595 | V995 Cyg | 11.882 | 9679 | 5705 | 0.04 | A |
| 9851944 | TYC 3558-939-1 | 11.249 | 7026 | 6902 | 1.24 | G1 |
| 9899416 | BR Cyg | 10.028 | 11056 | 6278 | 0.06 | A |
| 10156064 | TYC 3561-1283-1 | 10.367 | 7424 | 6268 | 0.07 | S |
| 10191056 | BD+47 2717 | 10.811 | 6588 | 6455 | 0.48 | S |
| 10206340 | V850 Cyg | 11.203 | 5844 | 4856[b] | 0.09 | A |
| 10486425 | 2MASS J19495442+4739323 | 12.465 | 7018 | 5847 | 0.08 | S |
| 10581918 | WX Dra | 12.796 | 8300 | 5544[b] | 0.05 | spec |
| 10619109 | TYC 3562-985-1 | 11.704 | 7028 | 3903 | $< 0.01$ | S |
| 10661783 | TYC 3547-2135-1 | 9.586 | 7764 | 6001 | $< 0.01$ | L |
| 10686876 | TYC 3562-961-1 | 11.727 | 7944 | 5842 | $< 0.01$ | S |
| 10736223 | V2290 Cyg | 13.621 | 7797 | 5069 | 0.01 | S |
| 10858720 | V753 Cyg | 10.971 | 7282 | 7223 | 0.90 | A |
| 12071006 | V379 Cyg | 13.533 | 7338 | 4660 | 0.01 | S |

[a] All flux ratios derived by maximizing the average cross-correlation functions over a grid of flux ratios unless otherwise noted (see Section 3.3 for details).

[b] Temperature ratio from Armstrong et al. (2014) used to determine secondary temperature

[c] Flux ratio adopted from Matson et al. (2016)

Note—Template temperature sources: A - Armstrong et al. (2014), G1 - Guo et al. (2016), G2 - Guo et al. (2017a), H - Hambleton et al. (2013), L - Lehmann et al. (2013), M - Matson et al. (2016), S - Slawson et al. (2011), spec - observed spectral type.

**Table 4**: Radial Velocity Measurements

| Object (KIC) | Date (HJD−2,400,000) | Orbital Phase[a] | $V_1$ (km s$^{-1}$) | $\sigma_1$ (km s$^{-1}$) | $(O-C)_1$ (km s$^{-1}$)[b] | $V_2$ (km s$^{-1}$) | $\sigma_2$ (km s$^{-1}$) | $(O-C)_2$ (km s$^{-1}$)[b] | Source (obs.) |
|---|---|---|---|---|---|---|---|---|---|
| 2305372 | 55367.8978 | 0.143 | −77.09 | 1.21 | −1.14 | 103.07 | 21.81 | −1.04 | KPNO |
| 2305372 | 56078.7768 | 0.218 | −92.72 | 1.07 | −1.16 | 126.70 | 21.69 | −8.42 | KPNO |
| 2305372 | 56079.8403 | 0.975 | 4.00 | 2.95 | 5.11 | 23.75 | 34.03 | ··· | KPNO |
| 2305372 | 56081.9033 | 0.444 | −39.07 | 1.25 | 2.02 | 46.51 | 22.33 | 5.24 | KPNO |
| 2305372 | 56486.8313 | 0.712 | 61.71 | 1.71 | −2.14 | −186.40 | 25.64 | −25.53 | KPNO |
| 2305372 | 56522.8387 | 0.346 | −77.17 | 1.78 | 2.06 | 95.86 | 25.48 | −17.31 | KPNO |
| 2708156 | 55368.8339 | 0.164 | −68.39 | 1.19 | 2.41 | 129.39 | 19.68 | −50.14 | KPNO |
| 2708156 | 55368.9349 | 0.217 | −74.35 | 1.24 | 3.02 | 39.07 | ··· | ··· | KPNO |
| 2708156 | 55718.8555 | 0.236 | −73.72 | 2.71 | 4.59 | −8.64 | ··· | ··· | Lowell |
| 2708156 | 55719.8914 | 0.784 | 25.45 | 2.86 | −2.75 | −256.29 | 31.55 | 1.12 | Lowell |
| 2708156 | 55720.9486 | 0.343 | −64.05 | 4.11 | 5.50 | −99.07 | ··· | ··· | Lowell |
| 2708156 | 55734.8661 | 0.702 | 29.56 | 1.64 | 2.58 | −249.80 | 21.87 | 1.92 | KPNO |
| 2708156 | 55734.9794 | 0.762 | 30.69 | 1.76 | 1.40 | −247.63 | 22.82 | 14.48 | KPNO |
| 2708156 | 55735.7758 | 0.183 | −70.87 | 1.45 | 2.91 | 313.26 | 20.61 | ··· | KPNO |
| 2708156 | 55735.9293 | 0.264 | −78.49 | 1.64 | −0.19 | 259.81 | 21.78 | 47.04 | KPNO |
| 2708156 | 55753.8980 | 0.765 | 51.57 | 3.05 | 22.37 | −239.01 | 33.73 | 22.72 | Lowell |
| 2708156 | 55754.7917 | 0.237 | −82.85 | 1.49 | −4.51 | 242.10 | 21.92 | 29.25 | Lowell |
| 2708156 | 55755.8035 | 0.772 | 28.26 | 1.50 | −0.64 | −247.51 | 22.22 | 12.94 | Lowell |
| 2708156 | 55755.9211 | 0.835 | 17.27 | 1.36 | −4.72 | −233.24 | 21.71 | −3.06 | Lowell |
| 2708156 | 56077.8059 | 0.030 | −55.33 | 1.34 | −20.74 | 38.86 | 19.50 | 20.04 | KPNO |
| 2708156 | 56077.8979 | 0.078 | −49.64 | 1.33 | 0.44 | 96.70 | 20.37 | 9.41 | KPNO |
| 2708156 | 56234.6275 | 0.949 | 2.90 | 1.39 | 10.31 | −98.16 | 20.76 | 3.15 | Lowell |
| 2708156 | 56522.8612 | 0.351 | −62.34 | 1.98 | 5.67 | 173.66 | 25.28 | 5.40 | KPNO |

[a] Relative to $T_0$ at primary eclipse.

[b] No data in $O-C$ columns indicates RV measurement was excluded from the orbital fit.

Note—Table 4 is available in its entirety in machine-readable format in the online journal. A portion is shown here for guidance regarding its form and content.

the cross-correlation function by fitting a linear slope to the relevant side of the background peak and subtracting out its contribution to the desired peak. The uncertainties in the secondary velocities were determined via the newly rectified peak in the same way as before, with occasional anomalously large errors due to tiny peaks replaced by errors determined via the method of Kurtz et al. (1992).

Once preliminary radial velocities were determined using an estimated flux ratio based on the ratio of the stellar radii and blackbody fluxes from the template temperatures, cross-correlations were repeated over a grid of flux ratios for each system. The average maximum correlation functions for each flux ratio were then plotted to find the interpolated maximum via the numerical derivative. The flux ratio corresponding to this peak, reported in Table 3, was used to perform a final set of cross-correlations and derive radial velocities for each system. In a few cases, the average maximum correlation function continued to increase for progressively smaller flux ratios approaching zero. We believe this reflects discrepancies in our adopted temperatures and the difficulty of measuring cross-correlation functions with varying slopes and backgrounds, rather than a true flux ratio of zero. For these systems we therefore list $f_2/f_1 < 0.01$ in Table 3 and use 0.01 for the derivation of radial velocities.

The radial velocities are presented in Table 4, which lists the *Kepler* Input Catalog (KIC) number, time of observation in heliocentric Julian date (HJD), orbital phase, radial velocities ($V$), uncertainties ($\sigma$), and observed minus calculated ($O-C$) residuals from the spectroscopic fit (§ 3.3) for both components in all 41 systems, as well as the observatory where the data were taken. Orbital phase is determined relative to $T_0$, taken to be the epoch of primary eclipse in Gies et al. (2015). Note the period of KIC 4848423 (3.0 d) is adopted from Gies et al. (2015) and is consistent with our velocity measurements and the revised *Kepler* Eclipsing Binary Catalog (Kirk et al. 2016), whereas it was originally included in Rowe et al. (2015) as a (false positive) transiting planet candidate with half the period.

### 3.2. *Comparison with TODCOR*

To confirm the accuracy and reliability of our double cross-correlation scheme we compared our derived radial velocities with those from TODCOR (Zucker & Mazeh 1994), which represents a reliable standard for velocity measurements. We used a version of TODCOR written in IDL by James Davenport[3]. This code produces the $R(s_1, s_2, \alpha)$ matrix for given values of the flux ratio $\alpha$ and Doppler shifts $s_1$ and $s_2$, then determines the radial velocities using the IDL DERIV procedure to find the local maximum position of $R(s_1, s_2, \alpha)$. We found that the local maximum may sit on a sloping background in some cases where the companion is faint, so we added an option to remove the background before finding the position of the maximum.

We first applied the IDL version of TODCOR to synthetic spectra formed by co-adding templates with known Doppler shifts and flux ratio. In every test case, TODCOR recovered the adopted radial velocities of the primary and secondary within the uncertainties, and our own double cross-correlation scheme also produced velocities that matched the known model values. Next we compared radial velocities derived from both TODCOR and our double cross-correlation method for KIC 5738698, and found that both sets agreed within the mutual uncertainties. Similar good agreement was found between TODCOR velocities and other systems in this work. Thus, we are confident that the radial velocities we measured using our double cross-correlation method are reliable and unhampered by systematic errors.

### 3.3. *Orbital Solutions*

Orbital elements for each star were determined using a nonlinear, least-squares fitting routine (Morbey & Brosterhus 1974). The periods ($P$) and epochs ($T_0$) were fixed to the values obtained from the eclipse timings of Gies et al. (2015), with the epoch corresponding to the time of primary eclipse. For the fitting procedure radial velocities were weighted by their uncertainties ($\propto 1/\sigma^2$) while velocities that were clear outliers (e.g., where the cross-correlation was unreliable due to blended peaks and/or extreme flux ratios) were omitted and are shown as radial velocities without $O-C$ values in Table 4.

In order to derive orbital parameters and optimize our observing time we concentrated on obtaining spectra during velocity extrema or quadrature phases to best constrain the orbits using fewer velocity measurements. Because of the resulting partial orbital coverage for many of the systems and the nature of short-period binaries, we used circular orbits to fit the velocities, with three exceptions: KIC 4544587, 4851217, and 8196180. The first system is a known 2.18 d period eccentric binary ($e = 0.275$) with tidally induced pulsations, strong apsidal motion, and self-excited pressure and gravity modes studied in detail by Hambleton et al. (2013). We therefore use the eccentricity ($e$) and argument of periastron ($\omega$) determined by Hambleton et al. when fitting the spectroscopic orbit of KIC 4544587. KIC 4851217 and 8196180 were identified as eccentric in Gies et al. (2015) and have separations between their primary and secondary eclipses that deviate from one-half the period (as in a circular system) by more than $\pm 0.005$ in orbital phase (Kirk et al. 2016). We use $e \sin\omega$ and $e \cos\omega$ as reported by Slawson et al. (2011) to determine $e$ and $\omega$ for KIC 4851217 and hold them fixed when determining the orbital solution, as they are not well constrained by our radial velocities. However, the values of $e$ and $\omega$ given in Slawson et al. for KIC 8196180 did not agree with our derived radial velocities. We therefore fit for $e$ and $\omega$ based on the offset between ($e \cos\omega$) and duration ($e \sin\omega$) of the two eclipses in the *Kepler* light curve using the method outlined in Matson et al. (2016), obtaining values of $e = 0.18$ and $\omega = 145°$, which were then fixed to derive the spectroscopic orbital elements.

Orbital parameters for each system, including the period ($P$), time of primary eclipse ($T_0$), velocity semi-amplitudes of the primary ($K_1$) and secondary ($K_2$), systemic velocities of the primary ($\gamma_1$) and secondary ($\gamma_2$), eccentricity ($e$), argument of periastron ($\omega$), root mean square of the primary ($\mathrm{rms}_1$) and secondary ($\mathrm{rms}_2$) velocity fits, the derived mass ratio ($q = M_2/M_1$), inclination ($i$), semi-major axis ($a$), and derived masses of the primary ($\mathrm{M}_1$) and secondary ($\mathrm{M}_2$) are given in Table 5. The inclination values are taken from Slawson et al. (2011) and were used to determine the semi-major axis (a), mass of the primary ($\mathrm{M}_1$) and mass of the secondary ($\mathrm{M}_2$) from our derived $a \sin i$ and $m \sin^3 i$ products, unless otherwise indicated in the table. Note, however, that the $\sin i$ values reported in Slawson et al. (2011) were determined via the artificial intelligence pipeline EBAI (Eclipsing Binaries via Artificial Intelligence; Prša et al. 2008), which determines approximate model parameters ($T_2/T_1$, $(R_1 + R_2)/a$, $e \sin\omega$, $e \cos\omega$, and $\sin i$) from eclipsing binary light curves. This process can sometimes determine unphysical model parameters, such as $\sin i > 1$, as listed for three of the systems examined here. For these systems (KIC 2305372, 10858720, and 12071006) we adopt an inclination of 90° (see Table 5).

[3] https://github.com/jradavenport/jradavenport_idl/blob/master/todcor.pro

**Table 5**: Orbital Solutions

| KIC | $P$ (d) | $T_0$ (HJD−2, 400, 000) | $K_1$ (km s$^{-1}$) | $K_2$ (km s$^{-1}$) | $\gamma_1$ (km s$^{-1}$) | $\gamma_2$ (km s$^{-1}$) | $e$ | $\omega$ (deg) | rms$_1$ (km s$^{-1}$) | rms$_2$ (km s$^{-1}$) | $M_2/M_1$ | $i$ (deg) | $a$ ($R_\odot$) | $M_1$ ($M_\odot$) | $M_2$ ($M_\odot$) |
|---|---|---|---|---|---|---|---|---|---|---|---|---|---|---|---|
| 2305372 | 1.40469145 | 55693.58496 | 80 ± 2 | 152 ± 7 | −14 ± 1 | −14[a] | 0.0 | ··· | 2.3 | 14.7 | 0.52 ± 0.03 | 90[g] | 6.4 ± 0.2 | 1.2 ± 0.1 | 0.62 ± 0.04 |
| 2708156 | 1.89126932 | 55690.03950 | 54 ± 2 | 238 ± 8 | −25 ± 2 | −25[a] | 0.0 | ··· | 8.3 | 25.3 | 0.23 ± 0.01 | 84.2 | 11.0 ± 0.3 | 4.05 ± 0.3 | 0.92 ± 0.07 |
| 3241619 | 1.70334707 | 55694.50030 | 92.5 ± 0.5 | 135 ± 3 | −47.1 ± 0.4 | −47[a] | 0.0 | ··· | 1.2 | 7.7 | 0.69 ± 0.02 | 84.0 | 7.7 ± 0.1 | 1.24 ± 0.04 | 0.86 ± 0.02 |
| 3327980 | 4.23102194 | 55699.074202 | 89 ± 2 | 102 ± 3 | 10 ± 2 | 10 ± 2 | 0.0 | ··· | 4.6 | 5.8 | 0.87 ± 0.03 | 85.1 | 16.0 ± 0.3 | 1.66 ± 0.07 | 1.44 ± 0.06 |
| 3440230 | 2.88110031 | 55690.39643 | 38 ± 1 | 151 ± 6 | −5.4 ± 0.9 | −5.4[a] | 0.0 | ··· | 2.8 | 14.2 | 0.25 ± 0.01 | 86.8 | 10.7 ± 0.3 | 1.6 ± 0.1 | 0.40 ± 0.03 |
| 4544587 | 2.18911430 | 55689.673044 | 114 ± 4 | 135 ± 5 | −24 ± 4 | −19 ± 5 | 0.28[b] | 329[b] | 12.8 | 16.0 | 0.84 ± 0.04 | 87.9[b] | 10.4 ± 0.3 | 1.69 ± 0.1 | 1.42 ± 0.09 |
| 4574310 | 1.30622004 | 55644.346499 | 42.6 ± 0.6 | 188 ± 4 | −44.2 ± 0.6 | −44.2[a] | 0.0 | ··· | 1.6 | 11.4 | 0.227 ± 0.006 | 82.5 | 6.0 ± 0.1 | 1.38 ± 0.06 | 0.31 ± 0.01 |
| 4660997 | 0.56256070 | 55654.43913 | 139 ± 7 | 184 ± 5 | −32 ± 5 | −32[a] | 0.0 | ··· | 13.5 | 12.2 | 0.76 ± 0.05 | 80.5 | 3.6 ± 0.1 | 1.16 ± 0.07 | 0.88 ± 0.07 |
| 4665989 | 2.248067589 | 55646.917282 | 100.2 ± 0.4 | 134 ± 4 | −26.8 ± 0.4 | −26.8[a] | 0.0 | ··· | 1.3 | 13.1 | 0.75 ± 0.02 | 82.5 | 10.5 ± 0.2 | 1.77 ± 0.09 | 1.32 ± 0.05 |
| 4848423 | 3.0036461 | 55928.7090 | 91 ± 2 | 103 ± 2 | 18 ± 2 | 17 ± 2 | 0.0 | ··· | 6.0 | 7.4 | 0.88 ± 0.03 | 87.3[f] | 11.6 ± 0.2 | 1.22 ± 0.05 | 1.08 ± 0.04 |
| 4851217 | 2.47028283 | 55643.10900 | 115 ± 2 | 107 ± 2 | −26 ± 2 | −24 ± 2 | 0.03[c] | 211[c] | 7.0 | 7.2 | 1.08 ± 0.03 | 77.8 | 11.1 ± 0.2 | 1.43 ± 0.05 | 1.55 ± 0.05 |
| 5444392 | 1.51952889 | 55688.84830 | 123.8 ± 0.7 | 122.3 ± 0.8 | −13.1 ± 0.6 | −12.9 ± 0.7 | 0.0 | ··· | 2.0 | 2.3 | 1.013 ± 0.008 | 86.3 | 7.41 ± 0.03 | 1.17 ± 0.01 | 1.19 ± 0.1 |
| 5513861 | 1.51020825 | 55625.52615 | 121 ± 1 | 138 ± 2 | 9 ± 1 | 9[a] | 0.0 | ··· | 4.2 | 7.8 | 0.88 ± 0.02 | 80.9 | 7.82 ± 0.08 | 1.50 ± 0.04 | 1.32 ± 0.03 |
| 5621294 | 0.938905233 | 55693.432093 | 57 ± 3 | 238 ± 20 | 19 ± 3 | 19[a] | 0.0 | ··· | 5.0 | 37.1 | 0.24 ± 0.02 | 75.3 | 5.7 ± 0.4 | 2.2 ± 0.4 | 0.54 ± 0.08 |
| 5738698 | 4.80877396 | 55692.334870 | 88.1 ± 0.6 | 92.8 ± 1 | 8.0 ± 0.6 | 8 ± 1 | 0.0 | ··· | 2.3 | 3.6 | 0.95 ± 0.01 | 86.3[d] | 17.2 ± 0.1 | 1.52 ± 0.03 | 1.44 ± 0.02 |
| 6206751 | 1.24534281 | 55683.783072 | 26.8 ± 0.5 | 203 ± 3 | −50.4 ± 0.4 | −56 ± 3 | 0.0 | ··· | 1.2 | 8.0 | 0.132 ± 0.003 | 76.3 | 5.81 ± 0.08 | 1.50 ± 0.05 | 0.198 ± 0.007 |
| 7368103 | 2.18251508 | 55698.67906 | 21 ± 1 | ··· | −21.8 ± 0.7 | ··· | 0.0 | ··· | 2.0 | ··· | ··· | 78.0 | ··· | ··· | ··· |
| 8196180 | 3.67165963 | 55695.751468 | 67 ± 2 | ··· | −11.7 ± 0.8 | ··· | 0.18 | 145 | 2.0 | ··· | ··· | 82.4 | ··· | ··· | ··· |
| 8262223 | 1.61301473 | 55692.217903 | 21.8 ± 0.5 | 203 ± 2 | 23.1 ± 0.4 | 21 ± 1 | 0.0 | ··· | 1.5 | 4.8 | 0.107 ± 0.002 | 75.2 | 7.42 ± 0.05 | 1.91 ± 0.03 | 0.205 ± 0.005 |
| 8552540 | 1.06193426 | 55692.5027 | 121 ± 1 | 153 ± 2 | −12 ± 1 | −12 ± 2 | 0.0 | ··· | 4.1 | 6.3 | 0.79 ± 0.01 | 80.7 | 5.83 ± 0.05 | 1.32 ± 0.03 | 1.04 ± 0.02 |
| 8553788 | 1.60616365 | 55690.62207 | 8.6 ± 0.7 | ··· | 27.9 ± 0.6 | ··· | 0.0 | ··· | 1.7 | ··· | ··· | 69.9 | ··· | ··· | ··· |
| 8823397 | 1.506503700 | 55686.100164 | 22.4 ± 0.9 | 221 ± 10 | −21.9 ± 0.7 | −23 ± 7 | 0.0 | ··· | 2.2 | 21.9 | 0.101 ± 0.006 | 82.1 | 7.3 ± 0.3 | 2.1 ± 0.2 | 0.21 ± 0.02 |
| 9159301 | 3.04477215 | 55693.13726 | 36 ± 1 | 147 ± 4 | 1.5 ± 0.9 | 1.5[a] | 0.0 | ··· | 3.3 | 11.0 | 0.25 ± 0.01 | 81.1 | 11.2 ± 0.2 | 1.61 ± 0.08 | 0.40 ± 0.02 |
| 9357275 | 1.588298214 | 55643.386640 | 81.4 ± 0.9 | ··· | −10.1 ± 0.8 | ··· | 0.0 | ··· | 2.5 | ··· | ··· | 74.9 | ··· | ··· | ··· |
| 9402652 | 1.07310422 | 55689.37059 | 144 ± 2 | 144 ± 1 | 14 ± 1 | 15 ± 1 | 0.0 | ··· | 4.5 | 3.9 | 1.0 ± 0.01 | 80.7 | 6.19 ± 0.04 | 1.39 ± 0.02 | 1.39 ± 0.03 |
| 9592855 | 1.21932483 | 55691.664073 | 144 ± 2 | 145 ± 7 | 8 ± 1 | 8[a] | 0.0 | ··· | 3.4 | 12.5 | 0.99 ± 0.05 | 74.1 | 7.3 ± 0.2 | 1.7 ± 0.1 | 1.72 ± 0.08 |
| 9602595 | 3.5565129 | 55642.26003 | 36 ± 1 | 177 ± 4 | −14 ± 1 | −14[a] | 0.0 | ··· | 2.7 | 13.2 | 0.202 ± 0.008 | 86.5 | 15.0 ± 0.3 | 3.0 ± 0.1 | 0.60 ± 0.03 |
| 9851944 | 2.16390178 | 55646.150110 | 125 ± 1 | 117 ± 2 | −3 ± 1 | −1 ± 1 | 0.0 | ··· | 5.1 | 5.4 | 1.07 ± 0.02 | 74.5[e] | 10.7 ± 0.1 | 1.70 ± 0.04 | 1.83 ± 0.04 |
| 9899416 | 1.332564228 | 55689.965570 | 108 ± 1 | 210 ± 3 | −15.9 ± 0.9 | −16 ± 3 | 0.0 | ··· | 3.9 | 12.0 | 0.516 ± 0.009 | 82.8 | 8.44 ± 0.08 | 3.00 ± 0.07 | 1.55 ± 0.03 |
| 10156064 | 4.85593645 | 55648.858147 | 77 ± 2 | 114 ± 4 | 4 ± 1 | 3 ± 3 | 0.0 | ··· | 4.8 | 9.4 | 0.67 ± 0.03 | 82.8 | 18.5 ± 0.4 | 2.1 ± 0.1 | 1.44 ± 0.08 |
| 10191056 | 2.42749484 | 55688.135138 | 100 ± 2 | 119 ± 2 | −20 ± 2 | −20 ± 2 | 0.0 | ··· | 6.8 | 6.9 | 0.83 ± 0.02 | 80.5 | 10.7 ± 0.2 | 1.50 ± 0.05 | 1.25 ± 0.04 |
| 10206340 | 4.564405 | 55710.0929 | 42 ± 2 | 137 ± 2 | −27 ± 1 | −28 ± 2 | 0.0 | ··· | 5.5 | 8.2 | 0.30 ± 0.01 | 82.0 | 16.3 ± 0.3 | 2.14 ± 0.07 | 0.65 ± 0.03 |
| 10486425[h] | 5.2748181 | 55681.8090 | 70.8 ± 0.9 | 98 ± 3 | 11.6 ± 0.8 | −6 ± 3 | 0.0 | ··· | 2.4 | 8.0 | 0.72 ± 0.03 | 82.7 | 17.8 ± 0.3 | 1.57 ± 0.08 | 1.13 ± 0.05 |
| 10581918 | 1.80186369 | 55744.790056 | 23.4 ± 0.8 | 191 ± 7 | −64.1 ± 0.7 | −64.1[a] | 0.0 | ··· | 2.3 | 19.4 | 0.130 ± 0.005 | 88.1 | 7.1 ± 0.1 | 1.30 ± 0.06 | 0.169 ± 0.009 |
| 10619109 | 2.04516645 | 55642.31502 | 33 ± 1 | 163 ± 22 | −22.8 ± 0.7 | −22.8[a] | 0.0 | ··· | 1.2 | 33.1 | 0.20 ± 0.03 | 73.4 | 8.3 ± 0.9 | 1.50 ± 0.4 | 0.31 ± 0.07 |
| 10661783 | 1.23136328 | 55692.541890 | 21 ± 1 | 247 ± 4 | −40 ± 1 | −40[a] | 0.0 | ··· | 4.7 | 17.5 | 0.083 ± 0.006 | 82.4 | 6.6 ± 0.1 | 2.33 ± 0.09 | 0.19 ± 0.01 |
| 10686876 | 2.61841548 | 55632.12313 | 67 ± 2 | ··· | −7.3 ± 0.9 | ··· | 0.0 | ··· | 3.0 | ··· | ··· | 79.9 | ··· | ··· | ··· |
| 10736223 | 1.105094315 | 55692.697641 | 46 ± 2 | 211 ± 8 | −2.4 ± 0.9 | −2.4[a] | 0.0 | ··· | 1.1 | 8.4 | 0.22 ± 0.01 | 89.2 | 5.6 ± 0.2 | 1.6 ± 0.1 | 0.35 ± 0.03 |
| 10858720 | 0.952377618 | 55692.951773 | 151 ± 3 | 159 ± 6 | −32 ± 3 | −30 ± 5 | 0.0 | ··· | 7.3 | 13.3 | 0.95 ± 0.04 | 90[g] | 5.8 ± 0.1 | 1.50 ± 0.08 | 1.43 ± 0.07 |
| 12071006 | 6.0960363 | 55693.45930 | 15 ± 4 | 166 ± 17 | −44 ± 2 | −44[a] | 0.0 | ··· | 4.5 | 21.7 | 0.09 ± 0.03 | 90[g] | 22 ± 2 | 3.4 ± 0.7 | 0.3 ± 0.1 |

[a] Fixed to systemic value of the primary.
[b] Fixed to value derived by Hambleton et al. 2013.
[c] Fixed to value derived by Slawson et al. 2011.
[d] Fixed to value derived by Matson et al. 2016.
[e] Fixed to value derived by Guo et al. 2016.
[f] Fixed to value obtained from preliminary light curve fit.
[g] Inclination set to 90° as $\sin i > 1$ in Slawson et al. 2011. See text for more details.
[h] Possible triple system (see §4.3.1).

NOTE—The period ($P$) and epoch ($T_0$ at primary eclipse) were fixed to values from Gies et al. (2015). Inclination values are from Slawson et al. (2011) and were used to determine the semi-major axis ($a$), mass of the primary ($M_1$) and mass of the secondary ($M_2$) unless otherwise noted.

**Table 6**: Single-Lined Binary Orbital Parameters

| KIC | $K_1$ (km s$^{-1}$) | $\gamma_1$ (km s$^{-1}$) | $e$ | $\omega$ (deg) | $a_1 \sin i$ ($R_\odot$) | $f(m)$ ($M_\odot$) |
|---|---|---|---|---|---|---|
| 7368103 | $21 \pm 1$ | $-21.8 \pm 0.7$ | 0.0 | ··· | 0.89±0.05 | $0.0020 \pm 0.0004$ |
| 8196180 | $67 \pm 2$ | $-11.7 \pm 0.8$ | 0.18 | 145 | 4.8±0.1 | 0.110±0.008 |
| 8553788 | $8.6 \pm 0.7$ | $27.9 \pm 0.6$ | 0.0 | ··· | 0.27±0.02 | 0.00011±0.00002 |
| 9357275 | $81.4 \pm 0.9$ | $-10.1 \pm 0.8$ | 0.0 | ··· | 2.55±0.03 | 0.089±0.003 |
| 10686876 | $67 \pm 2$ | $-7.3 \pm 0.9$ | 0.0 | ··· | 3.4±0.1 | 0.080±0.007 |

The primary and secondary radial velocities for each system were fit separately, providing consistency checks of the fits as well as our derived radial velocities. The resulting systemic velocities of the primary ($\gamma_1$) and secondary ($\gamma_2$) generally agree, however, small disparities (typically a few times the uncertainty) occur in some of the systems. These differences are likely due to mismatches between the observed spectra and model templates and to the small number of observations. For systems with discrepant systemic velocities, especially those with small mass and/or flux ratios where the secondary radial velocities were not as well constrained, the systemic velocities of the secondary ($\gamma_2$) were fixed to the value derived from the primary as noted in Table 5. To verify that fixing the systemic velocities in this way did not alter the orbital solutions we also fit spectroscopic orbits to the measured radial velocities using a Markov Chain Monte Carlo (MCMC) method that solved for a common systemic velocity while fitting for the primary and secondary orbits simultaneously. We derived the posterior orbital parameters using the MCMC method with a Gibbs sampler as implemented in the JAGS package (Just Another Gibbs Sampler; Plummer 2003). See Guo (2016) for more details. The resulting velocity semi-amplitudes and systemic velocities were identical to the ones fit separately within or just slightly outside the quoted uncertainties (see Table 5). Any differences in the semi-amplitudes tended to be in the secondaries where the radial velocities are not as well constrained. Similarly, there were no apparent offsets between the commonly derived systemic velocities and those derived separately or fixed to the value of the primary.

We note that KIC 4678873 was listed as an eclipsing binary by the ASAS and HATNET surveys and was included in the *Kepler* Eclipsing Binary Catalogs of Prša et al. (2011) and Slawson et al. (2011). However, subsequent analysis revealed it to be a 'false positive' in which the variations in its light curve are caused by a neighboring eclipsing binary, believed to be KIC 4678875, five arc seconds north and slightly fainter than KIC 4678873 (Rowe et al. 2015). The spectroscopic observations of KIC 4678873 show it to be a constant velocity star, and we therefore present the radial velocities we measured in Table 4 but omit the system from Table 5.

In addition, while both primary and secondary radial velocities were measured for KIC 10486425, initial measurements of the radial velocities led to unrealistically small estimates of the semi-amplitudes. Inspection of several deep and isolated spectral lines showed that the profiles at Doppler shift maxima had extensions towards zero velocity indicating the presence of a non-shifted spectral component. The radial velocities and orbital parameters listed in Tables 4 and 5 represent values for the primary and secondary components assuming an F-type tertiary that contributes 40% of the total flux. See Section 4.3.1 for more details.

## 4. DISCUSSION OF RADIAL VELOCITY RESULTS

### 4.1. *Single-Lined Spectroscopic Binaries*

For five of the 41 systems only the primary member of the binary was definitively detected in our spectra. In these cases the correlation peaks for the velocity separations were not prominent enough to yield reliable measurements of the secondary velocities. Small flux ratios and differences between the component spectra and their corresponding templates, as well as small velocity differences between the components and varying S/N of the observed spectra, can all contribute to difficulty in measuring the secondary velocities. The last two factors can even vary from one observation to another, resulting in reliable secondary velocities from one spectrum of an object but not from another (Mazeh et al. 2003).

In general, these five single-lined systems have late-A/early-F type primaries ($7000 - 8000$ K) with weak eclipses in the *Kepler* light curves, especially the secondary eclipses, and flux ratios $f_2/f_1 < 5\%$. We therefore present these systems as single-lined spectroscopic binaries (SB1), measuring orbital parameters based on the primary component. The velocity semi-amplitude of the primary ($K_1$), systemic velocity ($\gamma_1$), eccentricity ($e$), and longitude of periastron for the primary ($\omega$) are reproduced in Table 6 with the projected semi-major axis of the primary, $a_1 \sin i$, and the mass function, $f(m)$. The radial velocities, orbital solutions, and residuals of the SB1 systems are plotted in Figure 2.

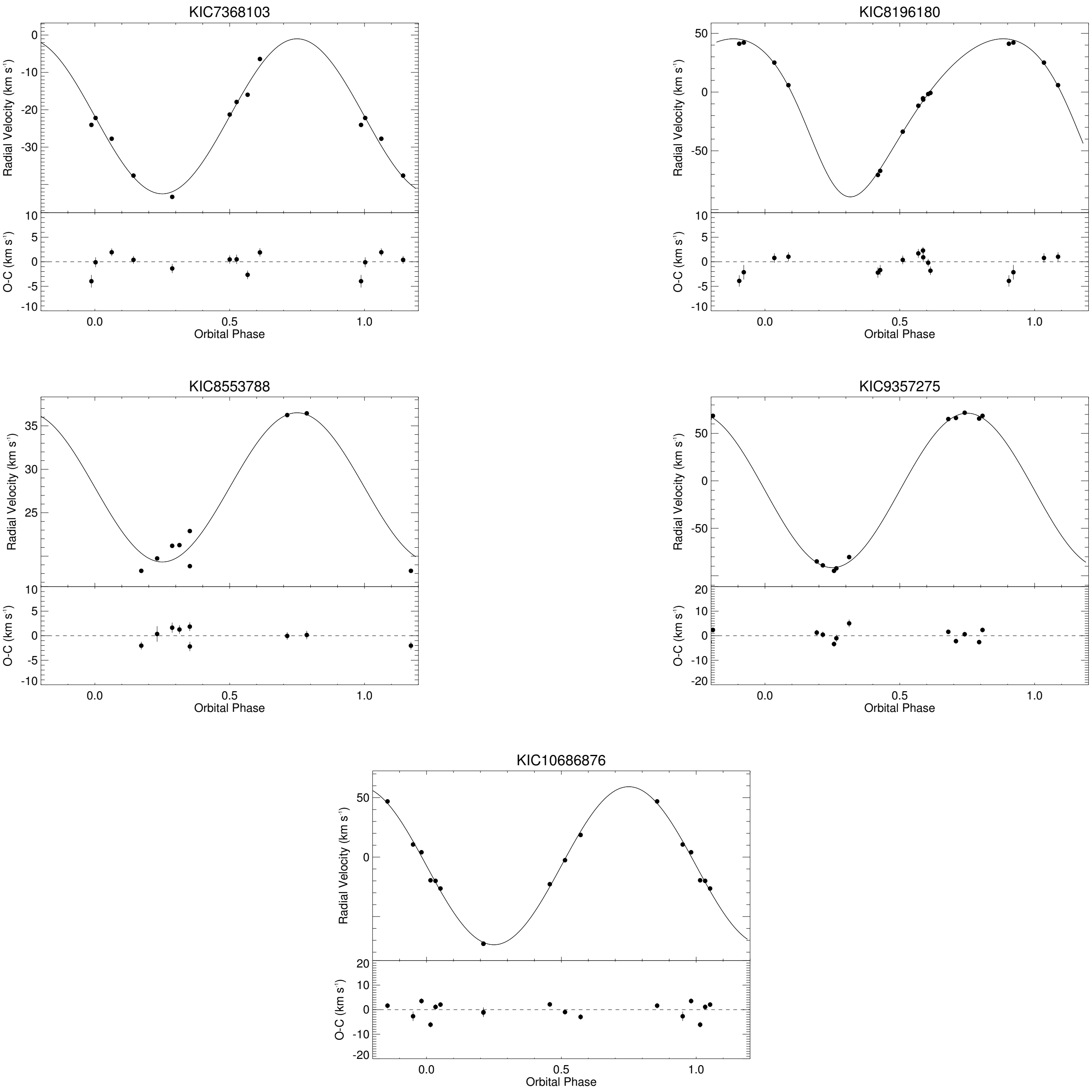

**Figure 2**: Radial velocities, spectroscopic orbits, and observed minus calculated $(O - C)$ values of SB1 systems.

### 4.2. *Double-Lined Spectroscopic Binaries*

Thirty-five of the remaining systems in our sample exhibited double lines in their spectra and/or were detected via cross-correlation allowing us to derive mass ratios ($q = M_2/M_1$) from the velocity semi-amplitudes of both stars. For the following discussion we assign the term "primary" (or star 1) to the hotter of the two components. The mass ratio distribution of binaries provides one of the few diagnostics for testing models of binary formation. While mass ratios are often determined for SB1 systems based on statistical techniques, the resolution ability of such techniques is limited, and they are most useful for examining general trends of the distribution (Mazeh et al. 2003). Therefore, dynamically determined mass ratios from double-lined spectroscopic binaries are valuable for deriving true mass ratio distributions and validating binary star formation scenarios. However, our sample suffers from severe selection effects (including our limited magnitude range, the brightness ratios imposed by the presence of eclipses, and using visual band spectra where the luminosity of stars less than $1\,M_\odot$ depends strongly on mass, impacting the detectability of companions) and does not provide a uniform sample suitable for statistical analysis. Having said that, the sample does allow us to examine trends in the derived mass ratios and compare them to previous results. Figure 3 shows a histogram of mass ratios for the 35 double-lined binaries divided into six bins: $0.0-0.16$, $0.17-0.33$, $0.34-0.50$, $0.51-0.66$, $0.67-0.83$, and $0.84-1.0$. As our sample is too small to draw general conclusions and combining the sample with previous studies could introduce biases, we use the mass ratio histogram to frame our discussion of binaries with similar properties and discuss whether the distribution appears to follow previously noted trends. The mass ratios in our sample fall into two distinct regions, with a peak at $q = 0.17-0.33$ and a second peak at $q = 0.84-1.0$. The portion of the plot with diagonal stripes represents systems with $q = M_2/M_1 > 1.0$, but are included in the plot as $q = M_1/M_2$. See Section 4.2.3 for more details.

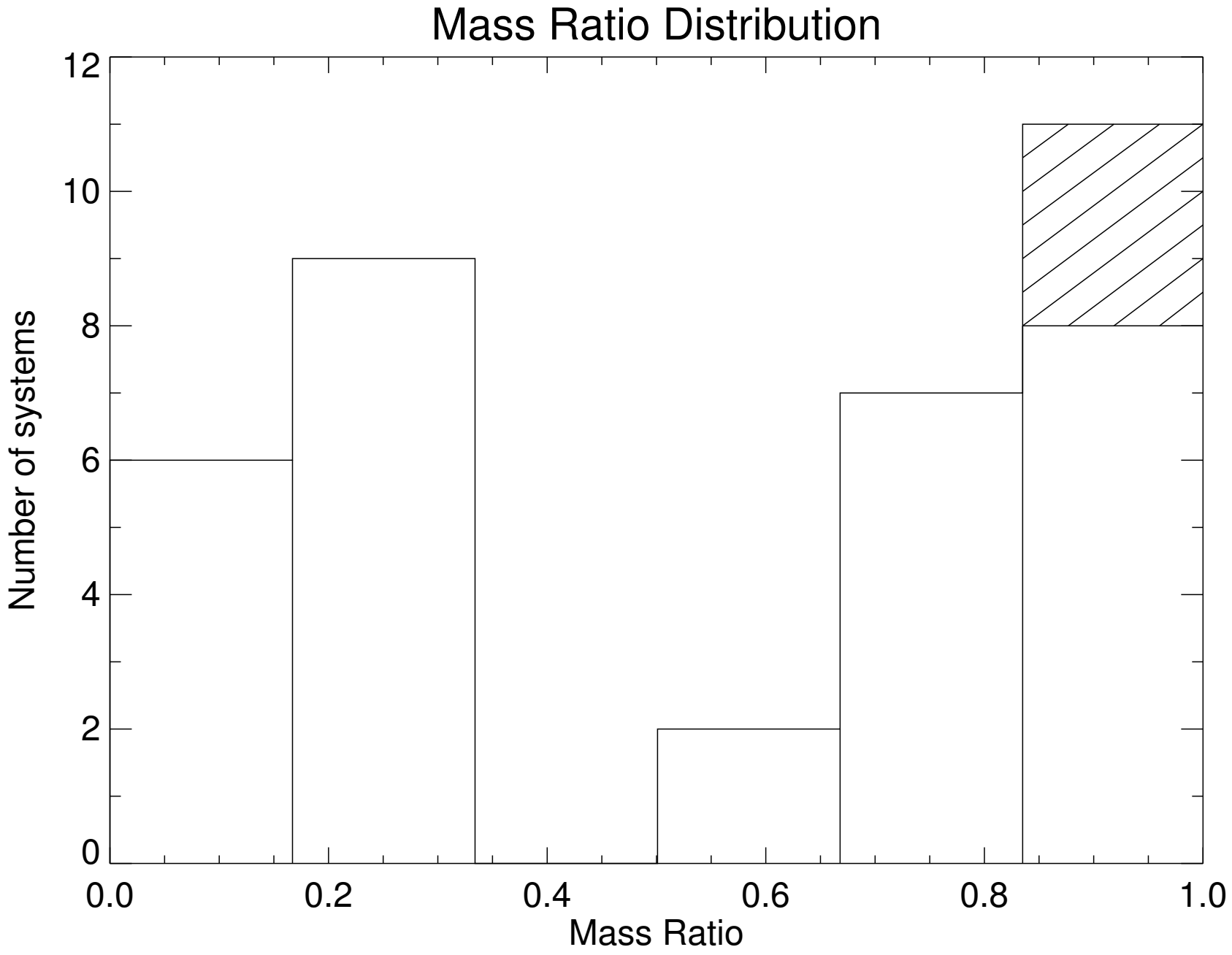

**Figure 3**: Mass ratio distribution for 34 double-lined spectroscopic binaries and the possible triple KIC 10484625 (see §4.3.1). The region with diagonal stripes represents systems with $q = M_2/M_1 > 1.0$, but are included in the plot as $q = M_1/M_2$ (see §4.2.3).

Studies of mass ratio distributions of spectroscopic binaries thus far have produced conflicting results with no consensus on the true mass ratio distribution. For example, Goldberg et al. (2003) examined 129 binaries (25 SB2s) with K-type primaries and periods between $1-2500$ days, finding a bimodal distribution similar to ours with a peak at $q \sim 0.2$ and a smaller peak at $q \sim 0.8$. In contrast, Gullikson et al. (2016) detected spectroscopic companions to 64 bright ($V < 6$) A and B-type stars and estimated their masses to infer a lognormal mass ratio distribution that peaks near $q \sim 0.3$. However, both Raghavan et al. (2010), who performed a comprehensive survey of companions to nearby solar-type stars, and Mazeh et al. (2003), who examined 62 (43 SB2s) main sequence and pre-MS binaries in the infrared to detect cooler companions, found relatively flat mass ratio distributions. Raghavan et al. reported a nearly flat distribution between $0.2 < q < 0.95$ with a strong peak at $q \sim 1$, demonstrating binaries, and in particular short-period systems, prefer like-mass pairs. Similarly, Mazeh et al. (2003) showed a flat distribution for $q > 0.3$, with an increase below $q = 0.3$ due to primarily long period systems. While such analyses have not produced a universal mass ratio distribution, several suggest separate distributions for long and short-period binaries (Duchêne & Kraus 2013) and a dependence on the mass of the primary, as massive binaries are known to favor companions of comparable mass while low-mass systems are more consistent with a flat mass-ratio distribution (Podsiadlowski 2014).

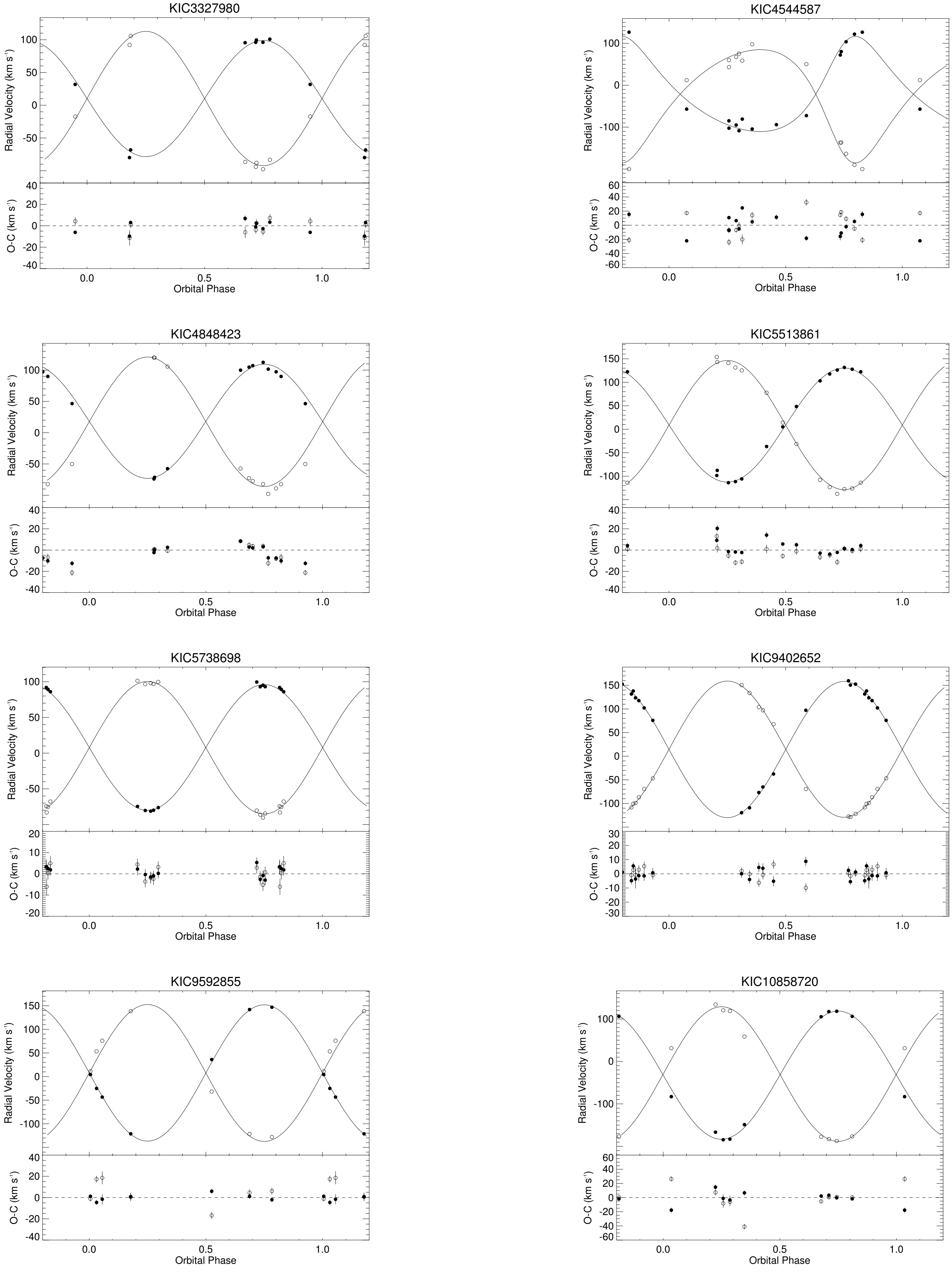

**Figure 4**: Radial velocities, spectroscopic orbits, and observed minus calculated $(O - C)$ values for SB2 systems with similar mass components $(0.84 \leq q \leq 1.0)$.

#### 4.2.1. *Similar Mass Binaries:* $0.84 \leq q \leq 1.0$

The peak at $0.84 \leq q \leq 1.0$ in our mass ratio distribution is roughly consistent with the studies of Goldberg et al. (2003) and Raghavan et al. (2010), which both found peaks for approximately like-mass binaries. There is ongoing debate about the existence of a population of 'twin' binaries based on an excess of systems with mass ratios from 0.95 to 1. While spectroscopic binaries with nearly identical components are found among all spectral types and in both long and short-period systems, a peak at $q \sim 1$ for solar-type short-period systems ($\sim 2 - 30$ d) has been found to be significant in some studies (e.g., Tokovinin 2000). Such like-mass pairs agree with theoretical simulations that show gas around proto-binaries preferentially accretes onto the secondary component, accumulating more mass until the components are roughly equal (Bate 1997).

In our sample of double-lined binaries only four systems have $0.95 \leq q \leq 1.0$ (KIC 5738698, 9402652, 9592855, and 10858720), not enough to affirm the 'twin' binary excess. However, there is a trend toward similar mass components, as nearly 25% of the SB2s have mass ratios between $0.84 - 1.0$, which increases to more than 30% when the $q = M_2/M_1 > 1.0$ systems (diagonal striped region of Fig. 3) are included. The radial velocities, orbital solutions, and residuals of the eight binary systems with similar mass components are plotted in Figure 4.

#### 4.2.2. *Intermediate mass secondaries:* $0.51 \leq q \leq 0.83$

Below the peak at $0.84 \leq q \leq 1.0$, the mass ratio distribution of our sample trails off until reaching the peak between $0.17 \leq q \leq 0.33$. The decreasing number of systems with lower mass ratios is consistent with the results of Goldberg et al. (2003), which show a decrease through $q = 0.5$ before increasing toward the peak at $q = 0.2$, but opposite those of Gullikson et al. (2016). The fewer systems observed with mass ratios between $0.51 - 0.83$ is also likely due to the increasing difficulty in detecting lower mass companions and more extreme flux ratios as well as the lack of any longer period systems ($P > 100$ d) that peak at lower mass ratios ($q \sim 0.2 - 0.3$) in the sample of Mazeh et al. (2003). The radial velocities, orbital solutions, and residuals of systems with mass ratios between $0.51 \leq q \leq 0.66$ and $0.67 \leq q \leq 0.83$ are plotted in Figures 5 and 6, respectively.

#### 4.2.3. *More massive secondaries:* $q > 1.0$

Three of the SB2 binaries have mass ratios greater than unity, in which the secondary component (the cooler star based on the weaker eclipse in the *Kepler* light curve) is more massive than the primary. Such systems are usually excluded from mass ratio distributions as they are typically evolved systems that have one or more components that differ from normal dwarf stars and, especially in short-period binaries, have experienced mass transfer that modifies the original mass ratio and no longer provides information about formation mechanisms. The radial velocities, orbital solutions, and residuals of the three systems where the secondary component is more massive than the primary are plotted in Figure 7.

KIC 9851944 ($q = 1.07 \pm 0.02$) was analyzed by Guo et al. (2016), who found that the components have very different radii (2.27 $R_\odot$, 3.19 $R_\odot$) despite their similar masses (1.76$M_\odot$, 1.79$M_\odot$) and temperatures (7026, 6920 K), indicating the hotter primary is still on the main sequence (MS) while the larger, cooler secondary has evolved to post-MS hydrogen shell burning. Using the more rigorous method described in Section 3.1 to derive updated radial velocities we find masses of 1.70 and 1.83$M_\odot$, which provide an even better match to the best-fit coeval MESA (Paxton et al. 2011, 2015) evolutionary models in Guo et al. (2016), confirming the evolutionary status of KIC 9851944.

KIC 5444392 was similarly found to have a mass ratio greater than one ($q = 1.013 \pm 0.008$). While the *Kepler* light curve shows minimal ellipsoidal variations (which can be indicative of an evolved star) the larger mass and lower temperature of the secondary suggests it may have evolved off the main sequence. The primary temperature (5965 K) is consistent with the derived mass of 1.17$M_\odot$ for a G0 main sequence star while $T_2 = 5725$ K is slightly too cool for $M_2 = 1.19\ M_\odot$ (Gray 2008), though this could be due to uncertainties in the temperature of the secondary. Analysis of the light curve and derivation of the component radii, which is beyond the scope of this work, is needed to confirm the evolutionary status of the system.

The final system with a mass ratio greater than unity is KIC 4851217, which has $q = 1.08 \pm 0.03$ and masses (1.43 and 1.55$M_\odot$) and temperatures consistent with mid F-type MS stars. However, the best-fit flux ratio determined by maximizing the correlation functions is $f_2/f_1 = 2.0$ indicating the secondary star gives off twice as much flux in the blue as the primary, while the light curve indicates the primary star is hotter. Thus, the cooler star has likely evolved within its Roche lobe to greater luminosity but a cooler temperature.

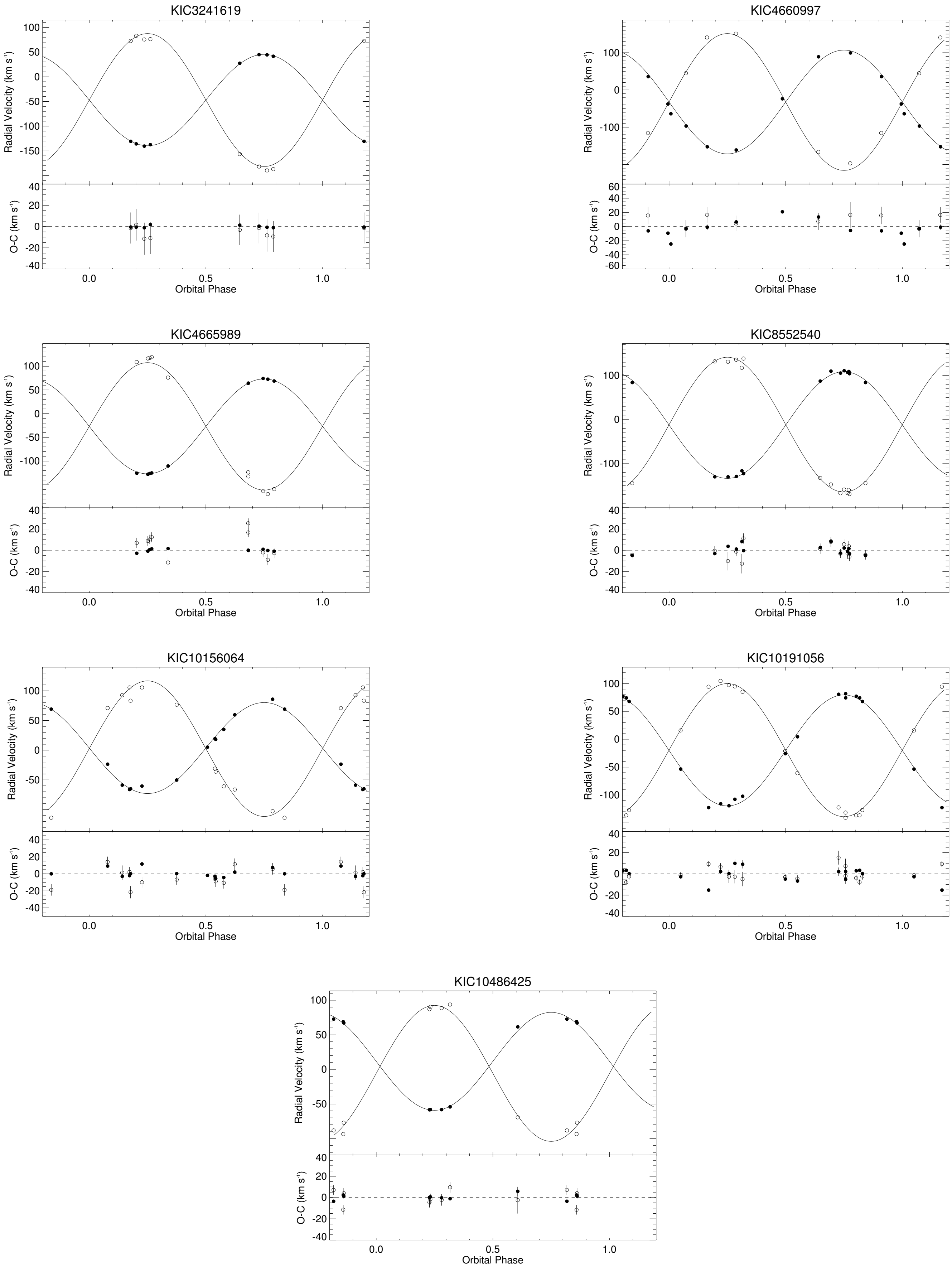

**Figure 5**: Radial velocities, spectroscopic orbits, and observed minus calculated $(O-C)$ values for SB2 systems with intermediate mass ratios $0.67 \leq q \leq 0.83$.

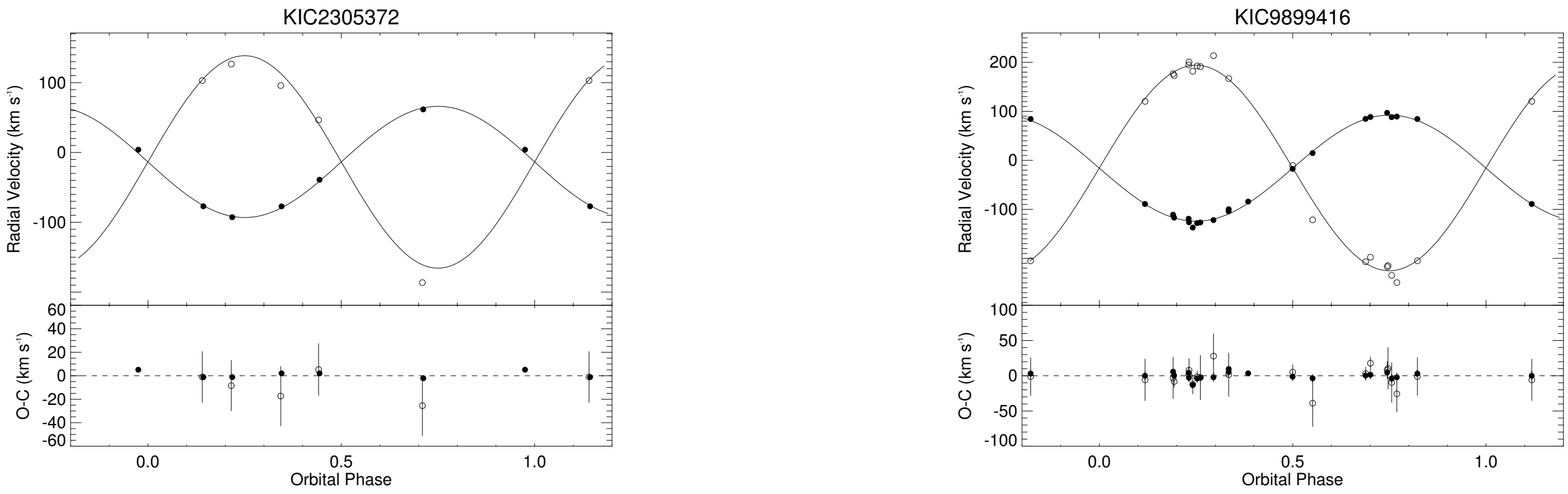

**Figure 6**: Radial velocities, spectroscopic orbits, and observed minus calculated $(O - C)$ values for SB2 systems with intermediate mass ratios $0.51 \leq q \leq 0.66$.

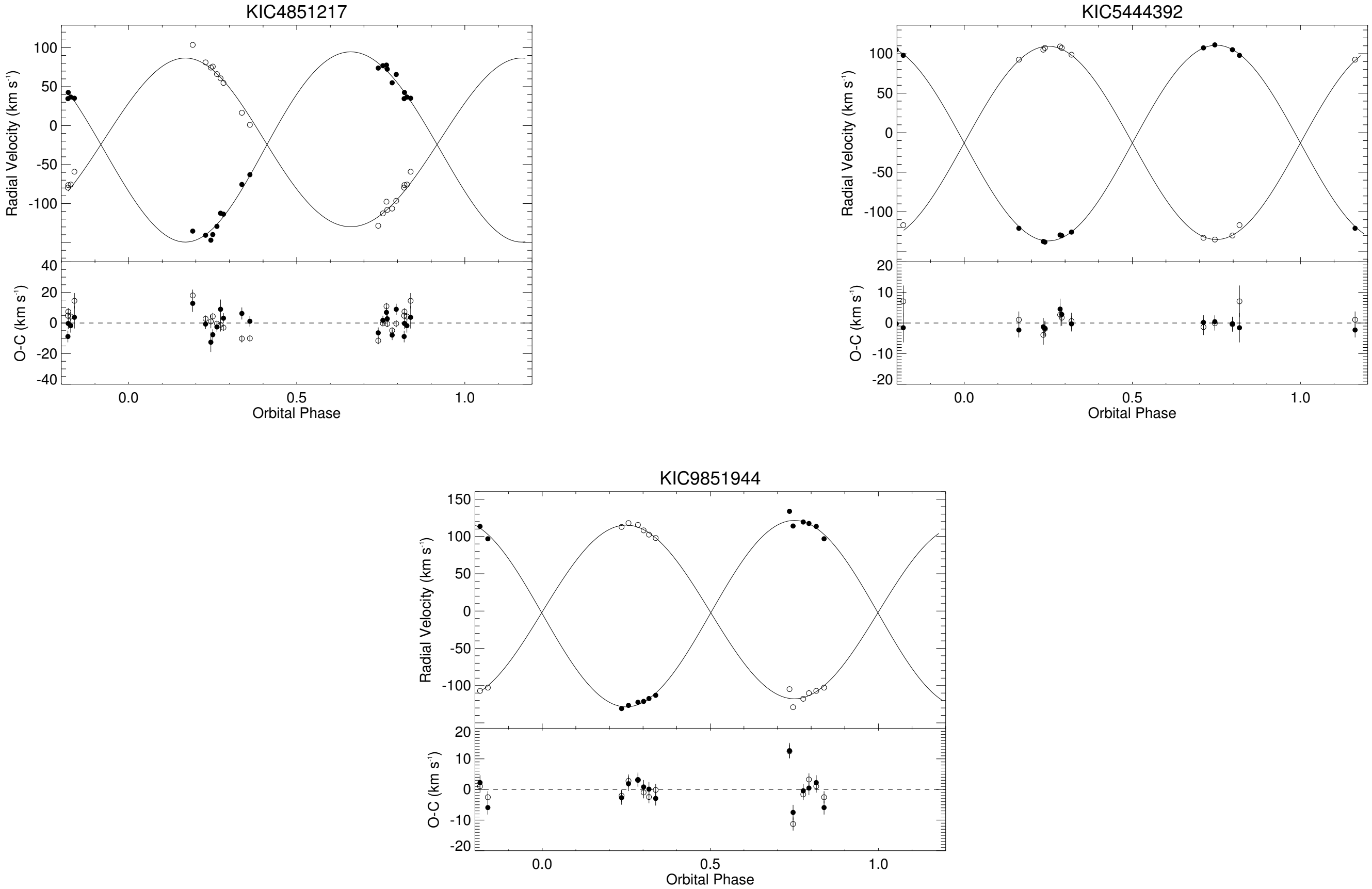

**Figure 7**: Radial velocities, spectroscopic orbits, and observed minus calculated $(O - C)$ values for SB2 systems with mass ratios greater than 1.0.

#### 4.2.4. *Algol-Type Binaries:* $q \leq 0.33$

The most dominant feature in the mass ratio distribution for the eclipsing and spectroscopic binaries we observed is the peak at $0.17 \leq q \leq 0.33$. While such a peak is seen in Goldberg et al. (2003) and Gullikson et al. (2016), the other studies discussed previously have reported a flat distribution or even a deficiency of low-mass companions. As we have focused on short-period systems that are thought to prefer like-mass companions, the presence of a low $q$ peak seems at odds with our expectations. However, when the secondary masses for systems with $q < 0.33$ are compared to the adopted temperatures, nearly all of the stars are hotter than expected for main sequence stars. This is highlighted in the plot of $\log M$ vs. $\log T_{\rm eff}$ in Figure 8, where plus signs represent the primary components and asterisks the secondary components in our sample. Yonsei-Yale isochrones (Demarque et al. 2004) for 0.1, 0.4, and 1.0 Gyr show the expected relationship between the mass and temperature of stars on the main sequence. Stars in detached binary (DB) systems that have not yet evolved or interacted should lie along the main sequence, and a significant portion of our sample do (approximately). A subset of binaries, however, display different $\log M - \log T$ relations for the primary and secondary components, with the secondaries having lower then expected masses for a given temperature, as expected for Algol-type binaries. Algols are semi-detached interacting binary (SDB) systems produced when the originally more massive component (mass loser) fills its Roche lobe and begins transferring mass to the less massive component (mass gainer) during central hydrogen burning (Case A) or after hydrogen in the core has been exhausted and shell hydrogen burning has begun (Case B). In general, these systems typically consist of an A or F-type primary with G or K-type subgiant or giant secondaries with masses of $0.2 - 0.4\ M_{\odot}$.

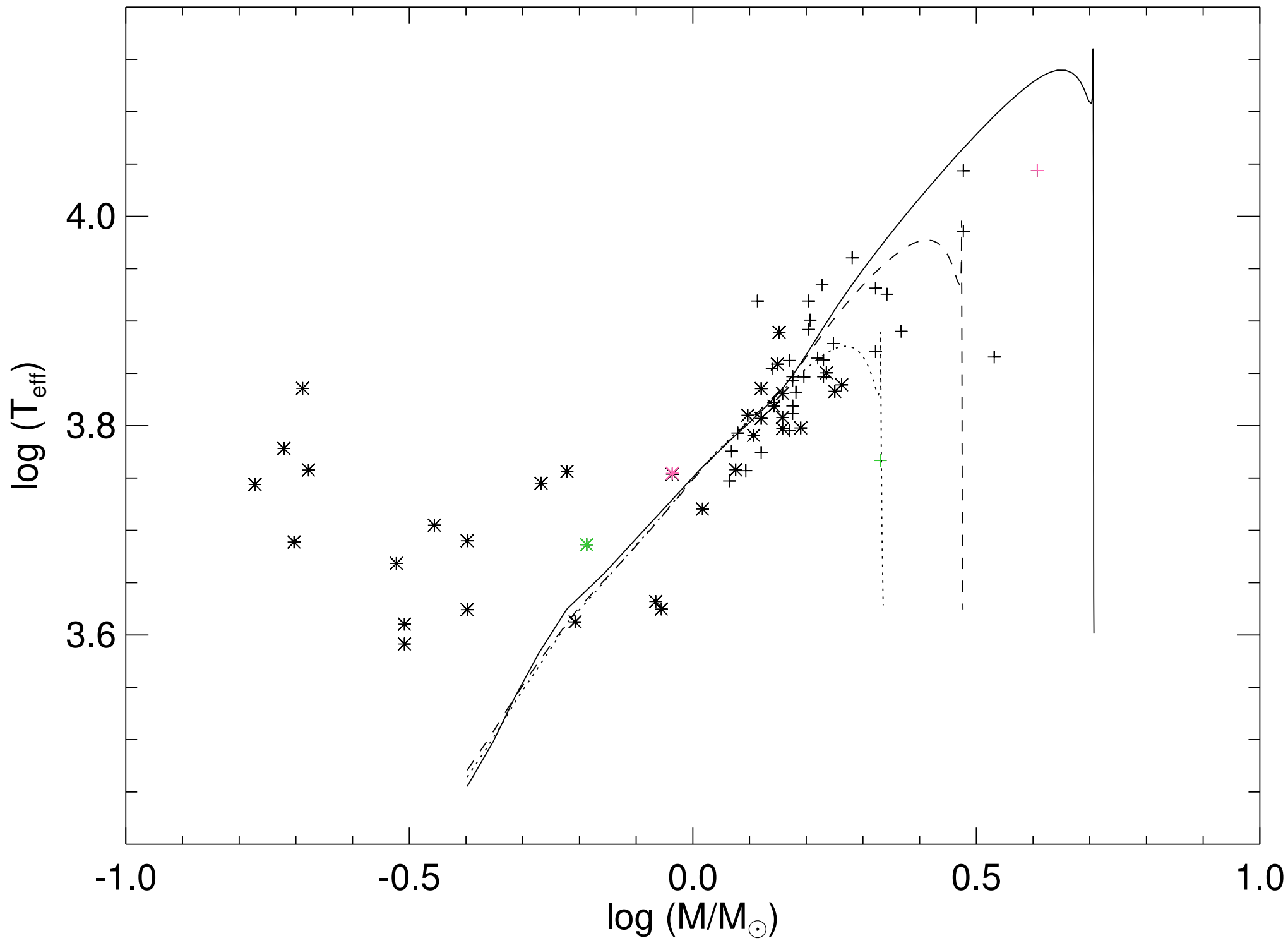

**Figure 8**: Locations of the primary (plus signs) and secondary (asterisks) components of the double-lined spectroscopic binaries in the $\log M - \log T$ plane. Yonsei-Yale ($Y^2$) Isochrones are plotted for 0.1 Gyr (solid line), 0.4 Gyr (dashed line), and 1.0 Gyr (dotted line). Pink and green symbols represent the components of KIC 2708156 and KIC 10206340, respectively; see text for more details.

The Algol candidates in our sample ($q \leq 0.33$) consist of primaries with masses between 1.3 and 4.05 $M_{\odot}$ corresponding to B, A, and F spectral types, with secondaries ranging from $0.169 - 0.92 M_{\odot}$ and a mean mass of $0.37 M_{\odot}$. All of the secondary components appear to the left of the main sequence in the $\log M - \log T$ plot (Fig. 8), with most significantly less massive then expected for main sequence stars. The two systems whose secondaries do not differ from the main sequence $\log M - \log T$ relation as much as the others and lie closest to the plotted main sequence are KIC 2708156 (pink symbols in Fig. 8), the most massive of the Algol systems in our sample ($M_1 = 4.05 M_{\odot}$, $M_2 = 0.92 M_{\odot}$), and KIC 10206340 (green symbols), which has the largest mass ratio ($q = 0.3$). However, as the primaries for both systems are notably to the right of the plotted main sequence, the temperatures may be underestimated for all components, in which case shifting the temperatures to higher values may place the secondaries more distinctly in the subgiant/giant region of the $\log M - \log T$ plane.

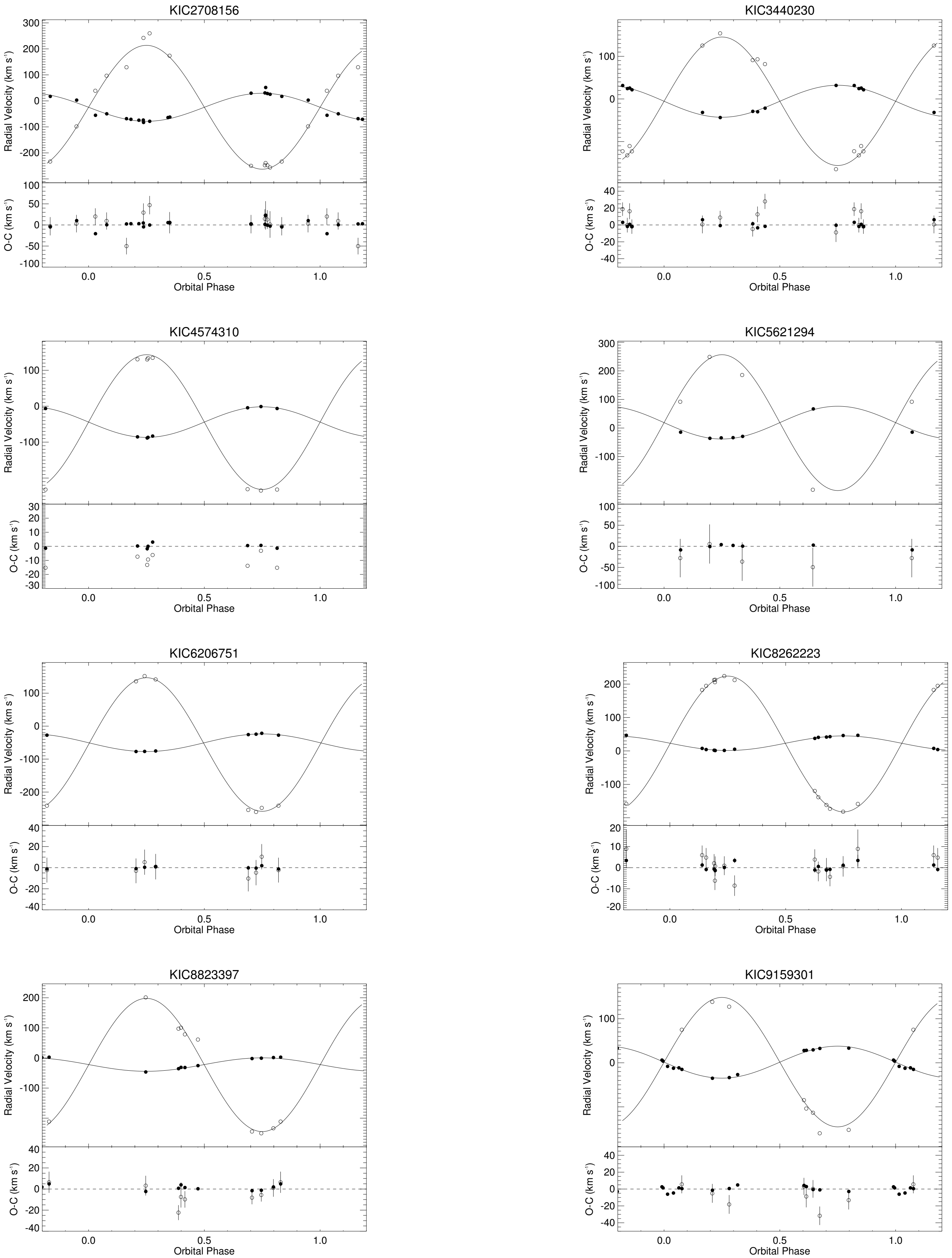

**Figure 9a**: Radial velocities, spectroscopic orbits, and observed minus calculated $(O - C)$ values for candidate Algol systems $(q < 0.33)$.

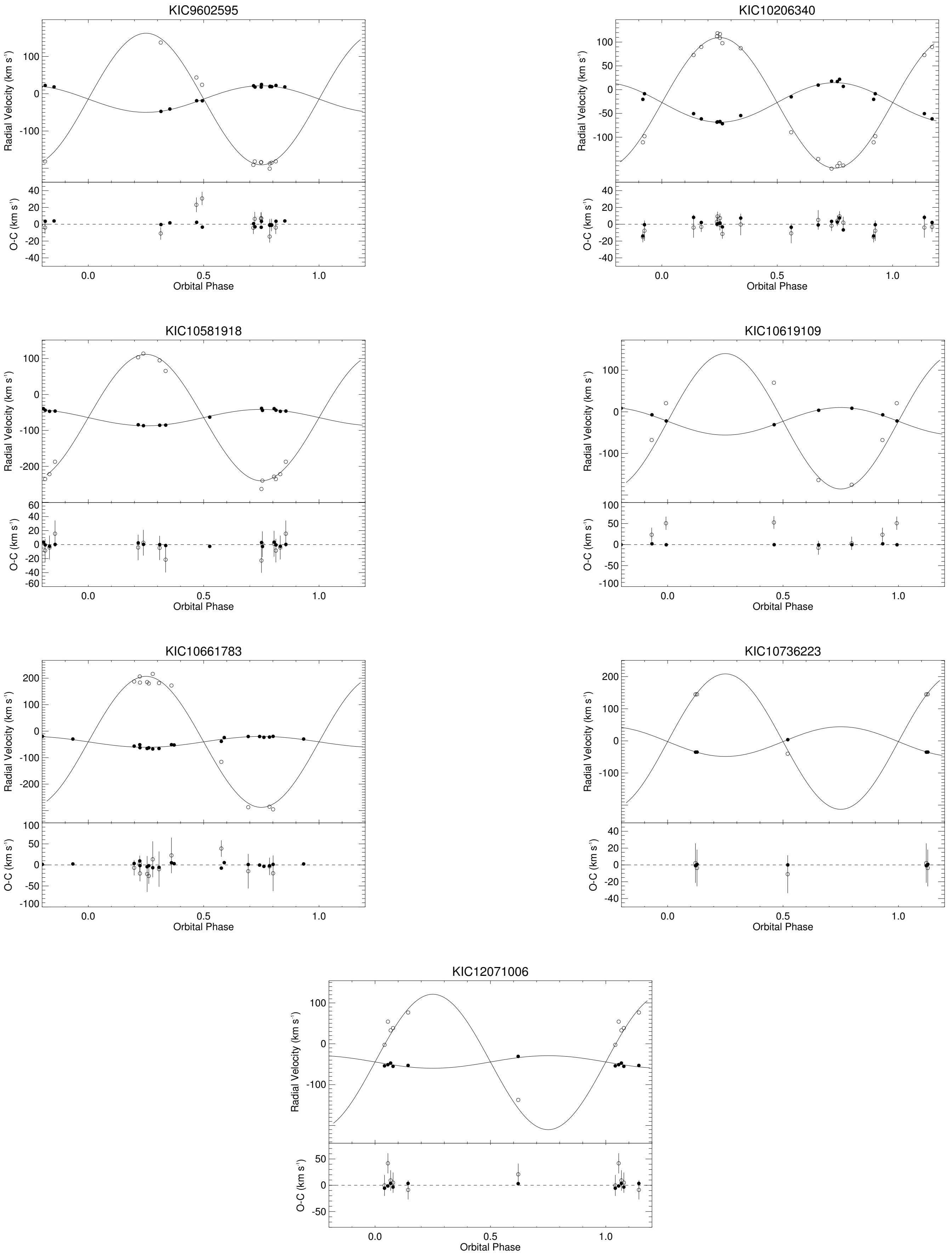

**Figure 9b**: Radial velocities, spectroscopic orbits, and observed minus calculated $(O - C)$ values for candidate Algol systems $(q \leq 0.33)$.

One of the systems, KIC 10661783, has been previously studied by Southworth et al. (2011) and Lehmann et al. (2013). Analysis of the *Kepler* light curve and spectroscopic observations revealed it to be a detached post-Algol binary system with $\delta$ Scuti pulsations in the primary. Lehmann et al. (2013) derived parameters of $M_1 = 2.10 \pm 0.03$, $R_1 = 2.58 \pm 0.02$, $T_1 = 7764 \pm 54$ for the primary and $M_2 = 0.191 \pm 0.003$, $R_2 = 1.12 \pm 0.02$, $T_2 = 6001 \pm 100$ for the secondary, with a slightly smaller primary mass then what we derive ($2.33 \pm 0.09$) but an equivalent secondary mass. The values for the secondary clearly confirm it has undergone mass loss in the past, though it has a short orbital period (1.23 d) and very low mass ratio ($q \sim 0.09$) compared to typical Algol systems. Based on comparisons with known Algol-type systems, Lehmann et al. (2013) determined KIC 10661783 to be a R CMa object, a small subgroup of stars characterized by the combination of a short period, low mass ratio, oversized secondary, and overluminous components. However, these systems are semi-detached binaries whereas Lehmann et al. (2013) were only able to fit the light curve and radial velocities of KIC 10661783 as a detached binary. In our sample, KIC 12071006 has similar temperatures and an equally small mass ratio ($q = 0.09$) but a longer period ($P = 6.10$ d). Further analysis of this system will provide an interesting comparison to KIC 10661783 and help inform our understanding of mass transfer and angular momentum redistribution in evolving binaries.

We have shown the $0.17 \leq q \leq 0.33$ peak in our mass ratio distribution is due to semi-detached Algol-type binaries, which are easier to detect due to their higher temperatures and increased luminosity, and is not a reflection of the true/original mass ratios of our sample. In a sample of 135 Algol-type eclipsing binaries with well determined parameters (74 detached (DB), 61 semi-detached (SDB) close binaries), Ibanoğlu et al. (2006) found more than 73% of the DBs had mass ratios larger than 0.80, with a mean value of $q = 0.88 \pm 0.14$, while the mass ratios of short-period ($< 5$ d) SDBs ranged from 0.11 to 0.57 with a mean of $q = 0.30$. This closely matches the mass ratio distribution of our sample, demonstrating the differences between evolved and unevolved systems and supporting the conjecture that unevolved systems have a preference for like-mass components. The radial velocities, orbital solutions, and residuals for the 15 candidate Algol systems are plotted in Figures 9a and 9b.

### 4.3. *Triple Star Systems*

Eclipse timing analysis of all 41 binaries in Gies et al. (2012, 2015) detected seven probable triple systems and long term trends indicative of a tertiary companion in seven additional systems. Preliminary orbital elements were determined for the seven systems that showed two inflection points in the $O - C$ changes for both the primary and secondary eclipses, including mass functions for the third star via (Mayer 1990)

$$f(m_3) = \frac{(m_3 \sin i)^3}{(m_1 + m_2 + m_3)^2} = \frac{1}{P_3^2}\left(\frac{173.15\,A}{\sqrt{1 - e^2\cos^2\omega}}\right)^3, \tag{1}$$

where $A$ is the observed semi-amplitude of the light travel time effect in days and $P_3$ is the period of the outer system in years. While we do not know the mass of the tertiary star nor the inclination of its orbit, we can further constrain its mass using the individual masses we derived for the primary and secondary components of the inner binary. The lower bound on the third star, $m_3 \sin i$, is calculated from the mass function for five systems with probable tertiaries in double-lined spectroscopic binaries. The primary and secondary masses derived from the spectroscopic orbits ($M_1, M_2$), the mass function ($f(m_3)$) of Gies et al. (2015), and $m_3 \sin i$ are shown in Table 7. The two additional systems with probable third components for which we were unable to detect confidently the secondary are listed for completeness.

Based on the $m_3 \sin i$ value calculated for KIC 2305372, the tertiary component is likely similar in mass or larger than the secondary component. We performed a three-star Doppler tomography reconstruction (Penny et al. 2001) in an attempt to detect the spectrum of the third star, but were unable to do so with any reliability. This is likely due to the extreme flux ratio of cool stars in the blue as well as the uncertainty in the mass of the third star (Borkovits et al. 2016, determine $m_3 \sin i = 0.41\,\mathrm{M}_\odot$). This system, however, would be an ideal target for high resolution spectroscopy, which may enable spectral reconstruction of all three stars in the future.

#### 4.3.1. *KIC 10486425*

In addition to the known and suspected triples found via eclipse timing, we believe one additional binary is a triple system due to the unrealistically small estimates of the semi-amplitudes from the radial velocity measurements of KIC 10486425, as mentioned previously. The presence of non-shifted spectral components in some deep and isolated lines during Doppler shift maxima led us to perform a three-star Doppler tomography reconstruction (Penny et al. 2001) of the individual spectrum of each star. The reconstructions revealed a mid F-type spectrum for the stationary component, with F0 V and G0 V types for the primary and secondary, respectively. Consequently, it appears that this is a spectroscopic triple system in which blending with the stationary lines of the tertiary reduces the absolute value of the velocity measurements of the primary and secondary components. We dealt with this by subtracting a model tertiary spectrum from the observed ones. The model was derived from the UVBLUE grid for $T_{\rm eff} = 6510$ K and $\log g = 4.3$ (representative main sequence values) that was shifted to the velocity of the tertiary in the reconstructed spectrum, $4.7 \pm 2.0$ km s$^{-1}$, and was rescaled to the expected flux contribution of the tertiary. We can only make a rough estimate of the tertiary's flux contribution in the observed spectral range, and unfortunately, this parameter has a large influence on the measured velocities. Subtracting the expected tertiary component removes absorption from the center of the observed profiles in such a way that the more that is removed, the weaker and more well separated the residual components from the primary and secondary appear. Thus, the larger the assumed tertiary contribution, the greater the absolute radial velocity measurements and the derived semi-amplitudes. We present here velocities for the primary and secondary (see Table 4) measured in difference spectra for a tertiary that contributes 40% of the total flux in the $B$-band covered by our spectra, but this is just one solution in a family based upon the adopted tertiary flux ratio. We encourage future higher resolving power spectroscopy of this system to resolve fully the three components and derive reliable semi-amplitudes. The radial velocities, orbital solution, and residuals for KIC 10486425 are plotted in Figure 5.

**Table 7**: Mass Constraints for Systems with Tertiaries

| KIC | $M_1$ ($M_\odot$) | $M_2$ ($M_\odot$) | $f(m_3)$ ($M_\odot$) | $m_3 \sin i$ ($M_\odot$) |
|---|---|---|---|---|
| 2305372 | $1.2 \pm 0.1$ | $0.62 \pm 0.04$ | $0.25 \pm 0.12$ | $0.9 \pm 0.2$ |
| 4574310 | $1.38 \pm 0.06$ | $0.31 \pm 0.01$ | $0.000032 \pm 0.000003$ | $0.045 \pm 0.002$ |
| 4848423 | $1.22 \pm 0.05$ | $1.08 \pm 0.04$ | $0.076 \pm 0.011$ | $0.74 \pm 0.04$ |
| 5513861 | $1.50 \pm 0.04$ | $1.32 \pm 0.03$ | $0.081 \pm 0.006$ | $0.86 \pm 0.02$ |
| 8553788 | ··· | ··· | $0.035 \pm 0.005$ | ··· |
| 9402652 | $1.39 \pm 0.02$ | $1.39 \pm 0.03$ | $0.0259 \pm 0.0006$ | $0.585 \pm 0.005$ |
| 10686876 | ··· | ··· | $0.019 \pm 0.005$ | ··· |

The *Kepler* light curve of KIC 10486425 was analyzed by Aliçavuş & Soydugan (2014), who used the first two quarters of data to obtain binary parameters and perform a frequency analysis. They derived two solutions, one with the eccentricity and third-light contribution fixed at zero and a second where these were left as free parameters. The second solution has larger values of $T_2$ (5727 vs. 5210 in the first solution) and $q$ (0.59 vs. 0.40), which are (slightly) more consistent with our derived parameters, but the third light contribution was still found to be zero. This disagreement with our results will only be resolved through future high angular resolution and high spectral resolving power observations.

#### 4.3.2. *KIC 10191056*

One of the stars in our sample, KIC 10191056, has been reported elsewhere as a visual binary, which may indicate the system is also a triple. Couteau (1983) observed a separation of 1.2 arcsec for KIC 10191056 and Ziegler et al. (2017) measured a separation of 1.32 arcsec and a $\Delta$ mag of 1.90 at 6000Å (via Robo-AO) and 1.54 in Kp band (via NIRI at Gemini-N). In addition, recent work by Hełminiak et al. (2017) detected three sets of lines visible in the spectra of KIC 10191056 using high-resolution echelle spectra ($R = \lambda/\delta\lambda \approx 50,000$) in the range $4360 - 7535$Å. They assumed the visual companion is gravitationally bound to the inner, eclipsing binary and responsible for the third set of (narrow) spectral features to derive $M_{A,a} = 1.590 \pm 0.032\, M_\odot$, $M_{A,b} = 1.427 \pm 0.036\, M_\odot$, and $M_B = 0.16 \pm 1.34\, M_\odot$, where $Aa + Ab$ refers to the inner binary and $B$ is the outer component. However, they note that because the change of the systemic velocity of the inner pair is indistinguishable from zero the outer component may not be gravitationally bound or has a period too large to be seen in the radial velocities, implying the third set of spectral lines may be due to another body.

In our analysis of KIC 10191056 we did not detect any evidence of an additional component in the spectra, nor were we able to recover the spectra of a tertiary through three-star Doppler tomography. As Hełminiak et al. (2017) note, however, the rotationally broadened features of the eclipsing pair affect the precision of the radial velocity measurements and our spectral resolution is considerably lower than theirs. Our derived velocity semi-amplitudes and mass ratio agree with those of Hełminiak et al. within $\lesssim 3\sigma$, though their analysis incorporates both radial velocities and *Kepler* light curves and includes a third body. Similar agreement is found for KIC 8552540, another star analyzed by both us and Hełminiak et al. but without a detected tertiary companion.

## 5. SUMMARY

Spectroscopic observations of the 41 eclipsing binaries in our sample have resulted in 454 spectra that were used to measure radial velocities through cross-correlation with template spectra. One of the stars, KIC 4678873, is shown to be a constant velocity star and not an eclipsing binary. Radial velocities were measured for only the primary component in five of the systems, resulting in orbital parameters based on the primary, including $a_1 \sin i$ and the mass function $f(m)$. In systems where velocities were measured for the primary and secondary we derived orbital elements for both components and found values of $m_1 \sin^3 i$, $m_2 \sin^3 i$, and $a \sin i$. Using inclinations from light curve fits by Slawson et al. (2011), we determine masses and semi-major axis values for 35 double-lined spectroscopic binaries. We analyze the resulting mass ratio distribution, identifying 15 semi-detached Algol systems that have undergone Roche lobe overflow and mass transfer. Three additional systems show evolved secondaries, while the remaining systems appear to be unevolved. The mass ratio distribution also demonstrates the tendency for short-period binaries to have similar mass components, likely a result of gas preferentially accreting onto the lower-mass component until reaching comparable masses during formation (Bate 1997).

For five of the seven systems with eclipse timing variations indicative of a third body we use our derived masses for the primary and secondary to determine minimum masses of the tertiary. Four of the distant companions may be K-type stars with $0.5\,M_\odot < M_3 < 0.9\,M_\odot$, while the fifth may be substellar ($M_3 \geq 0.045\,M_\odot$). We detect what is likely another triple system via spectroscopy, as KIC 10486425 had unrealistically small semi-amplitudes and indications of unshifted spectral features during Doppler shift maxima. We derive masses of $1.57\,M_\odot$ and $1.13\,M_\odot$ for the primary and secondary stars by subtracting a mid F-type model tertiary spectra shifted to the velocity of the third star.

We acknowledge the observations taken using the 4 m Mayall telescope at KPNO and are grateful to the director and staff of KPNO for their help in obtaining these observations. We also thank the anonymous referee for providing constructive comments that improved the paper. *Kepler* was competitively selected as the tenth Discovery mission. Funding for this mission is provided by NASA's Science Mission Directorate. This study was supported by NASA grants NNX12AC81G, NNX13AC21G, and NNX13AC20G. This material is based upon work supported by the National Science Foundation under Grant No. AST-1009080 and AST-1411654. Institutional support has been provided from the GSU College of Arts and Sciences and from the Research Program Enhancement fund of the Board of Regents of the University System of Georgia, administered through the GSU Office of the Vice President for Research and Economic Development.

*Facility:* Kepler, Mayall, Perkins, DAO:1.85m.

*Software:* IRAF, PROCOR (Gies & Bolton 1986), ATLAS9, SYNTHE, TOCDOR (Zucker & Mazeh 1994), todcor.pro (`https://github.com/jradavenport/jradavenport_idl/blob/master/todcor.pro`)

## REFERENCES

Aliçavuş, F. K., & Soydugan, E. 2014, in IAU Symposium, Vol. 301, Precision Asteroseismology, ed. J. A. Guzik, W. J. Chaplin, G. Handler, & A. Pigulski (Cambridge, UK: Cambridge University Press), 433

Armstrong, D. J., Gómez Maqueo Chew, Y., Faedi, F., & Pollacco, D. 2014, MNRAS, 437, 3473

Bate, M. R. 1997, MNRAS, 285, 16

Borkovits, T., Hajdu, T., Sztakovics, J., et al. 2016, MNRAS, 455, 4136

Borucki, W. J., Koch, D., Basri, G., et al. 2010, Science, 327, 977

Cenarro, A. J., Peletier, R. F., Sánchez-Blázquez, P., et al. 2007, MNRAS, 374, 664

Conroy, K. E., Prša, A., Stassun, K. G., et al. 2014, AJ, 147, 45

Couteau, P. 1983, A&AS, 53, 441

Demarque, P., Woo, J.-H., Kim, Y.-C., & Yi, S. K. 2004, ApJS, 155, 667

Duchêne, G., & Kraus, A. 2013, ARA&A, 51, 269

Eggleton, P. P., Kisseleva-Eggleton, L., & Dearborn, X. 2007, in IAU Symposium, Vol. 240, Binary Stars as Critical Tools & Tests in Contemporary Astrophysics, ed. W. I. Hartkopf, P. Harmanec, & E. F. Guinan, 347–355

Fabrycky, D., & Tremaine, S. 2007, ApJ, 669, 1298

Famaey, B., Jorissen, A., Luri, X., et al. 2005, A&A, 430, 165

Gies, D. R., & Bolton, C. T. 1986, ApJS, 61, 419

Gies, D. R., Matson, R. A., Guo, Z., et al. 2015, AJ, 150, 178

Gies, D. R., Williams, S. J., Matson, R. A., et al. 2012, AJ, 143, 137

Goldberg, D., Mazeh, T., & Latham, D. W. 2003, ApJ, 591, 397

Gray, D. F. 2008, The Observation and Analysis of Stellar Photospheres (Cambridge, UK: Cambridge University Press)

Gullikson, K., Kraus, A., & Dodson-Robinson, S. 2016, AJ, 152, 40

Guo, Z. 2016, PhD thesis, Georgia State University, Atlanta, GA

Guo, Z., Gies, D. R., & Matson, R. A. 2017b, ApJ, submitted, arXiv:1704.03789

Guo, Z., Gies, D. R., Matson, R. A., & García Hernández, A. 2016, ApJ, 826, 69

Guo, Z., Gies, D. R., Matson, R. A., et al. 2017a, ApJ, 837, 114

Hambleton, K. M., Kurtz, D. W., Prša, A., et al. 2013, MNRAS, 434, 925

Hartman, J. D., Bakos, G., Stanek, K. Z., & Noyes, R. W. 2004, AJ, 128, 1761

Hełminiak, K. G., Ukita, N., Kambe, E., et al. 2017, ArXiv e-prints, arXiv:1702.03311

Ibanoǧlu, C., Soydugan, F., Soydugan, E., & Dervişoǧlu, A. 2006, MNRAS, 373, 435

Kirk, B., Conroy, K., Prša, A., et al. 2016, AJ, 151, 68

Kurtz, M. J., Mink, D. J., Wyatt, W. F., et al. 1992, in ASP Conf., Vol. 25, Astronomical Data Analysis Software and Systems I, ed. D. M. Worrall, C. Biemesderfer, & J. Barnes (San Francisco, CA: ASP), 432

Latham, D. W., & Stefanik, R. P. 1992, in IAU Transactions, Vol. XXI B, XXIst General Assembly - Transactions of the IAU, ed. E. J. Bergeron (Dordrecht: Kluwer Academic Publishers), 269

Lehmann, H., Southworth, J., Tkachenko, A., & Pavlovski, K. 2013, A&A, 557, A79

Luck, R. E., & Heiter, U. 2006, AJ, 131, 3069

Massarotti, A., Latham, D. W., Stefanik, R. P., & Fogel, J. 2008, AJ, 135, 209

Matson, R. A., Gies, D. R., Guo, Z., & Orosz, J. A. 2016, AJ, 151, 139

Mayer, P. 1990, Bulletin of the Astronomical Institutes of Czechoslovakia, 41, 231

Mazeh, T., Simon, M., Prato, L., Markus, B., & Zucker, S. 2003, ApJ, 599, 1344

Molenda-Zakowicz, J., Frasca, A., Latham, D. W., & Jerzykiewicz, M. 2007, AcA, 57, 301

Morbey, C. L., & Brosterhus, E. B. 1974, PASP, 86, 455

Naoz, S. 2016, ARA&A, 54, 441

Naoz, S., Farr, W. M., & Rasio, F. A. 2012, ApJL, 754, L36

Nidever, D. L., Marcy, G. W., Butler, R. P., Fischer, D. A., & Vogt, S. S. 2002, ApJS, 141, 503

Paxton, B., Bildsten, L., Dotter, A., et al. 2011, ApJS, 192, 3

Paxton, B., Marchant, P., Schwab, J., et al. 2015, ApJS, 220, 15

Penny, L. R., Seyle, D., Gies, D. R., et al. 2001, ApJ, 548, 889

Petrie, R. M., Andrews, D. H., & Scarfe, C. D. 1967, in IAU Symposium, Vol. 30, Determination of Radial Velocities and their Applications, ed. A. H. Batten & J. F. Heard (London, UK: Academic Press), 221

Pigulski, A., Pojmański, G., Pilecki, B., & Szczygieł, D. M. 2009, AcA, 59, 33

Plummer, M. 2003, in DSC Working Papers, Proceedings of the 3rd International Workshop on Distributed Statistical Computing, ed. K. Hornik, F. Leisch, & A. Zeileis (Wien, Vienna: Austria: Technische Universitat)

Podsiadlowski, P. 2014, in Accretion Processes in Astrophysics, ed. I. González Martínez-País, T. Shahbaz, & J. Casares Velázquez, XXI Canary Islands Winter School of Astrophysics (Cambridge, UK: Cambridge University Press), 45

Prša, A., Guinan, E. F., Devinney, E. J., et al. 2008, ApJ, 687, 542

Prša, A., Batalha, N., Slawson, R. W., et al. 2011, AJ, 141, 83

Raghavan, D., McAlister, H. A., Henry, T. J., et al. 2010, ApJS, 190, 1

Rappaport, S., Deck, K., Levine, A., et al. 2013, ApJ, 768, 33

Rodríguez-Merino, L. H., Chavez, M., Bertone, E., & Buzzoni, A. 2005, ApJ, 626, 411

Rowe, J. F., Coughlin, J. L., Antoci, V., et al. 2015, ApJS, 217, 16

Slawson, R. W., Prša, A., Welsh, W. F., et al. 2011, AJ, 142, 160

Southworth, J., Zima, W., Aerts, C., et al. 2011, MNRAS, 414, 2413

Tokovinin, A. 2014a, AJ, 147, 86

—. 2014b, AJ, 147, 87

Tokovinin, A., Thomas, S., Sterzik, M., & Udry, S. 2006, A&A, 450, 681

Tokovinin, A. A. 2000, A&A, 360, 997

Valenti, J. A., & Fischer, D. A. 2005, ApJS, 159, 141

Ziegler, C., Law, N. M., Morton, T., et al. 2017, AJ, 153, 66

Zucker, S. 2003, MNRAS, 342, 1291

Zucker, S., & Mazeh, T. 1994, ApJ, 420, 806